\documentclass[reprint,aps,superscriptaddress,amsmath,amssymb,longbibliography,nofootinbib]{revtex4-1}
\usepackage{graphicx}
\usepackage{dcolumn}
\usepackage{bm}
\usepackage{color}
\usepackage{xspace}
\usepackage{ulem}
\usepackage{xfrac}
\usepackage[colorlinks=true,urlcolor=blue,citecolor=blue,linkcolor=blue,bookmarks=false,pdfstartview={FitH}]{hyperref}

\def \cm-1{cm$^{-1}$\,}

\begin{document} 
\title{Lattice dynamics and spin-phonon coupling in the kagome spin ice HoAgGe}                            

\author{Shangfei Wu}
\email{wusf@baqis.ac.cn}
\affiliation{Beijing Academy of Quantum Information Sciences, Beijing 100193, China}

\author{Lingxiao Zhao}
\affiliation{Department of Physics, Southern University of Science and Technology, Shenzhen, China}

\author{Wei Song}
\affiliation{Department of Physics, Southern University of Science and Technology, Shenzhen, China}

\author{Mingshu Tan}
\affiliation{School of Physical Science and Technology, Lanzhou University, Lanzhou 730000, China}

\author{Feng Jin}
\affiliation{Beijing National Laboratory for Condensed Matter Physics, Institute of Physics, Chinese Academy of Sciences, Beijing 100190, China}

\author{Tianping Ying}
\affiliation{Beijing National Laboratory for Condensed Matter Physics, Institute of Physics, Chinese Academy of Sciences, Beijing 100190, China}

\author{Jia-Xin Yin}
\affiliation{Department of Physics, Southern University of Science and Technology, Shenzhen, China}

\author{Qingming Zhang}
\email{qmzhang@iphy.ac.cn}
\affiliation{School of Physical Science and Technology, Lanzhou University, Lanzhou 730000, China}
\affiliation{Beijing National Laboratory for Condensed Matter Physics, Institute of Physics, Chinese Academy of Sciences, Beijing 100190, China}


\date{\today}               
                                                                                                                                                                                                              
\begin{abstract}  
     
We employ polarization-resolved Raman spectroscopy and first-principles phonon calculations to study the lattice dynamics and spin-phonon coupling in the kagome spin ice compound HoAgGe. 
Upon cooling, HoAgGe shows transitions from a nonmagnetic state at 300\,K to a partially magnetic ordered phase below $T_2=11.6$\,K, eventually reaching a fully magnetic ordered phase below $T_1=7$\,K. We detect 8 of 10 Raman-active phonon modes at room temperature, with frequencies consistent with first-principles phonon calculations. We find that the low-energy phonon modes harden upon cooling,  soften slightly below $T_2$, and, finally harden again below $T_1$, suggesting finite spin-phonon coupling in HoAgGe. 
Furthermore, one high-energy mode at 185\,\cm-1 with $E'$ symmetry exhibits a Fano lineshape due to electron-phonon coupling.
Our Raman results establish finite spin-phonon coupling and electron-phonon coupling in the magnetic phase of HoAgGe, providing a foundation for future studies of the novel field-induced phases at low temperatures in HoAgGe.
                                                                                                                                               
\end{abstract}                  
                                                                                                                                                                                      
                                                                                                                                                                                                                                                                                                                                                                                        
\maketitle

\section{INTRODUCTION}\label{INTRODUCTION}

The kagome lattice is a model system to study electronic and magnetic properties~\cite{Itiro1951}. Especially, the corner-shared triangle network of the kagome lattice is an ideal platform to study geometrical frustration, which leads to novel phenomena such as quantum spin liquids, spin ice, and macroscopic degeneracy in ground states without long-range magnetic ordering~\cite{Gilbert_2016_PhysicsToday,Balents2010,Broholm_2020_Science}. 
Spin ice, whose local moments replicate the frustration seen in the arrangement of hydrogen ions in frozen water, has been a focus of modern condensed matter physics. The study of these materials has introduced a variety of fascinating concepts of significant interest. At low energies, these materials exhibit an emergent gauge field, and their excitations manifest as magnetic monopoles, which result from the fractionalization of the microscopic spin degrees of freedom~\cite{Steven_2001_Science,Castelnovo_2012_review,Castelnovo2008,Bramwell_2020}. 
Khomskii proposed that in spin ice compounds, magnetic monopoles should be accompanied by electric dipoles~\cite{Khomskii_2010,Khomskii2012}. The association of electric dipoles with magnetic monopoles allows for the study and manipulation of these exotic monopoles using electric fields, thus providing a way for practical applications~\cite{Khomskii2012,Grams_2014_NC}.
This proposal was confirmed in the quantum spin ice candidate Tb$_2$Ti$_2$O$_7$, which hosts giant spin-lattice (spin-phonon) coupling~\cite{aleksandrov1985crystal,Ruff_2010_PhysRevLett,Klekovkina_2011,Fennell_2014_PhysRevLett,Ruminy_2019_PhysRevB} in a Raman scattering experiment~\cite{Jin_2020_PhysRevLett}.

\begin{figure*}[!ht] 
\begin{center}
\includegraphics[width=2\columnwidth]{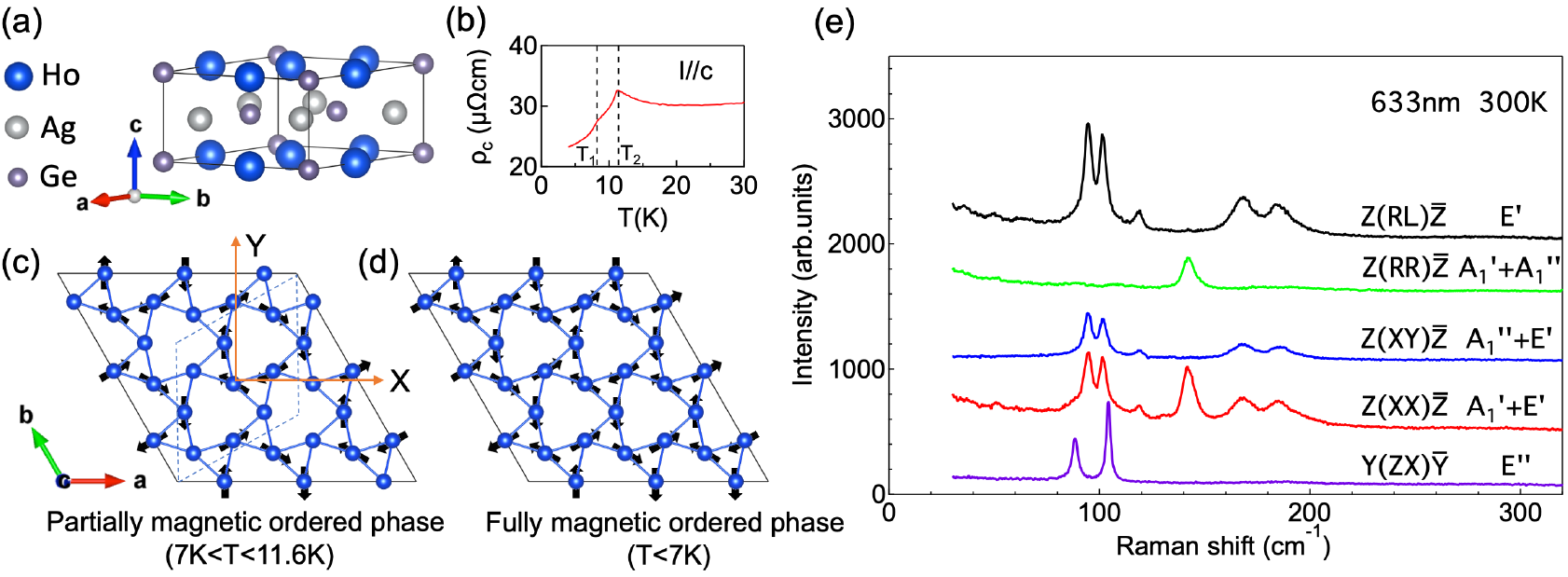}
\end{center}
\caption{\label{Fig1_300K} 
(a) Crystal structure of HoAgGe. (b) $T$ dependence of $c$-axis resistivity $\rho_{c}$ of HoAgGe. The dashed lines mark the partially magnetic ordered phase transition at $T_2=11.6$\,K and the fully magnetic ordered phase transition at $T_1=7$\,K. (c) and (d) Illustrations of the magnetic structure for HoAgGe in the partially magnetic ordered phase ($T_1<T<T_2$) and fully magnetic ordered phase ($T<T_1$), respectively. Ag and Ge atoms are omitted for simplification. The black arrows denote the spin directions, the dashed lines in (c) denote the $\sqrt{3} \times \sqrt{3}$ magnetic unit cell, and the orange arrows in  (c) denote the definitions of the $X$ and $Y$ directions.
(e) Raman spectra in the $XX$, $XY$, $RR$, $RL$, and $ZX$ scattering geometries at 300\,K recorded by a 633\,nm laser line.}
\end{figure*}

Recently, a kagome spin ice compound exhibiting a fully ordered ground state was established in HoAgGe~\cite{Zhao_2020_science}. HoAgGe is part of the $R$AgGe ($R$=Tb$-$Lu) series, characterized by ZrNiAl-type intermetallics with a noncentrosymmetric $P\bar{6}2m$ space group.
In this structure, rare-earth ions form a distorted two-dimensional kagome lattice of corner-sharing triangles~[Fig.~\ref{Fig1_300K}(a)]~\cite{GIBSON199634,BARAN_1998}. 
Each Ho$^{3+}$ atom in HoAgGe has ten $4f$ electrons with a ground state of $^{5}I_8$. From the Curie-Weiss fit of the magnetic susceptibility data, an effective magnetic moment m$_\text{eff}$ = 10.6\,$\mu_\text{B}$ is reported. The system adheres to two-dimensional ice rules defined for in-plane Ising-like classical spins on the kagome lattice, which results in local configurations of two-in$-$one-out or one-in$-$two-out spin arrangements on the triangles. 
The spin ice state evolves 
from a nonmagnetic state at room temperature to a partially magnetic ordered state at $T_2=11.6$\,K~[Fig.~\ref{Fig1_300K}(c)], and finally to a fully magnetic ordered phase at $T_1=7$\,K [Fig.~\ref{Fig1_300K}(d)]. 
These transitions are also visible in the $c$-axis resistivity measurement, with a sharp drop at $T_2$ and a kinklike feature at $T_1$~[Fig.~\ref{Fig1_300K}(b)]~\cite{Zhao_2020_science,Li_2022_PhysRevB}. With the magnetic field along the kagome plane,
it is noted that the magnetic structures at magnetization plateaus can be derived from the ground state by reversing certain Ho$^{3+}$ spins with Ising-like anisotropy while still following the ice rule~\cite{Zhao_2020_science}.  At these magnetization plateaus, it is reported that vanishing hysteresis exists in the field dependence of the magnetization at low temperatures but finite hysteresis exists in the field-dependent anomalous Hall conductivity~\cite{Zhao_2024_Nature_Phy}. Furthermore, at low temperatures in the spin ice state, scanning tunneling microscope measurements revealed a pair of pronounced dips in the local tunneling spectrum at symmetrical bias voltages with negative intensity values, indicating an inelastic tunneling signal linked to spin ice magnetism~\cite{Deng_2024PhysRevLett}. 
The Ising-like Ho$^{3+}$ spin reversals induced by the in-plane magnetic field 
potentially generate magnetic monopole and antimonopole pairs, along with the emergent electric dipoles attached to these magnetic monopoles~\cite{Khomskii_2010,Khomskii2012}. One important ingredient for such phenomena is to have spin-lattice coupling or spin-phonon coupling in this system. However, the lattice dynamics and the estimate of the spin-phonon coupling across these peculiar magnetic phase transitions in HoAgGe are not available in the current literature.
                  
In this paper, we use polarization-resolved Raman spectroscopy and first-principles phonon calculations to investigate the lattice dynamics and spin-phonon coupling in the kagome spin ice compound HoAgGe.  We detect 8 out of 10 Raman-active phonon modes at room temperature. The experimental phonon frequencies are consistent with first-principles phonon calculations. As the temperature drops, the low-energy phonon modes harden due to anharmonic phonon decay, soften slightly below $T_2$, and, finally, harden again below $T_1$, indicating the presence of finite spin-phonon coupling in HoAgGe. Additionally, an $E'$-symmetry mode at 185\,\cm-1 shows a Fano line shape due to finite electron-phonon coupling. The Raman results  in the present study confirm the existence of finite spin-phonon and electron-phonon coupling in the magnetic phase of HoAgGe, encouraging further research into the novel field-induced phases at low temperatures in this material.

\begin{figure*}[t] 
\begin{center}
\includegraphics[width=2\columnwidth]{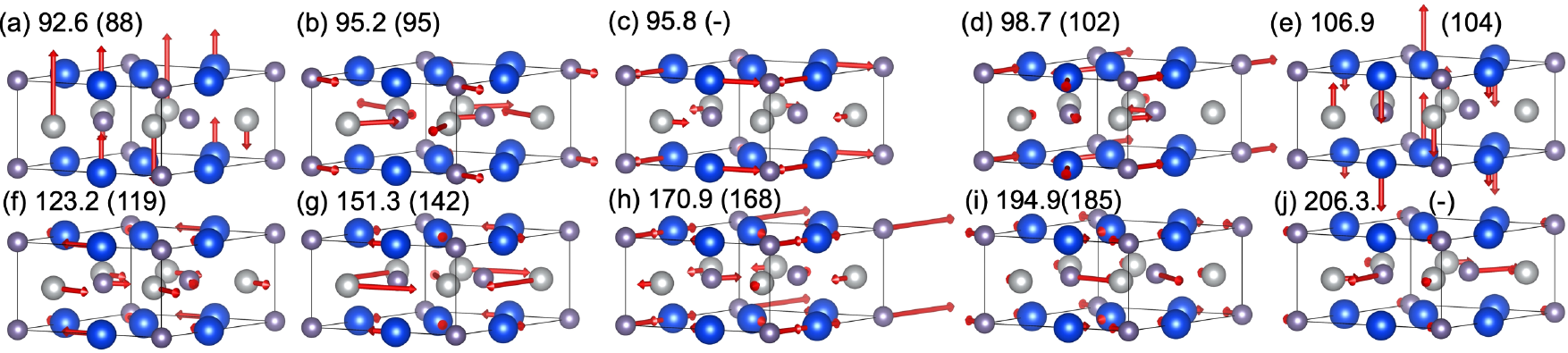}
\end{center}
\caption{\label{Fig2_pattern} 
Major lattice vibration patterns for the Raman-active modes of HoAgGe. The DFT calculated mode frequencies for each mode are shown in each panel while the experimental values are shown in the parentheses. The units are in  \cm-1.
}
\end{figure*}  
    
\section{Methods}\label{Methods}
                                                    
\textit{Single-crystal preparation and characterization.}\label{Crystal_preparation}                                                                                                            
Single crystals of HoAgGe were grown by using the Ag–Ge-rich self-flux method as described in detail in Ref.~\cite{Zhao_2020_science}. These samples were characterized by electric transport and magnetic susceptibility measurements at zero magnetic field.
The extracted fully magnetic ordered phase transition temperature is $T_1= 7$\,K, and the partially magnetic ordered phase transition temperature is $T_2=11.6$\,K. 

The as-grown HoAgGe single crystal was sliced and meticulously polished to expose its (100) and (001) crystallographic planes.
A strain-free area for Raman scattering measurement was examined
by a polarized optical image. The strain-free area was further examined by comparing the phonon
linewidths obtained on the as-grown (001) crystallographic plane and the polished (001) crystallographic plane. We did not find any noticeable polishing-induced linewidth broadening.
                                                                                         
\textit{Raman scattering measurements.}\label{Raman}                                          
Raman-scattering experiments were performed with a Horiba Jobin-Yvon LabRAM HR evolution spectrometer. One notch filter and two Bragg filters were used in the collection optical path to clean the laser line from the backscattered light.
The polished HoAgGe sample was positioned in a closed-cycle helium-gas-exchange cryostat (AttoDRY 2100), which allows for cooling down to the base temperature of 1.8\,K.
The Raman measurements were mainly performed using a HeNe laser line at 632.8\,nm (1.96\,eV) in
a backscattering geometry. 
The excitation laser beam was focused onto a spot 5\,$\mu$m in diameter, with an incident power less than 0.3\,mW. 
Linear and circular polarizations were used in this study to distinguish the symmetry of the Raman modes.
Spectra were recorded with a 1800 mm$^{-1}$ grating and liquid-nitrogen-cooled CCD detector.
The instrumental resolution was maintained at better than 0.75\,\cm-1 (full width at half maximum).
All linewidth data presented were corrected for the instrumental resolution. 
The temperatures shown in this paper were corrected for laser heating (see Appendix~\ref{laser_heating_determination}).

                                                                                                                                                                                                        
The Raman spectra were recorded from the $ab$ (001) and (100) surface for scattering geometries denoted as $\mu v = XX, XY, RR, RL$, which is short for $Z(\mu v)\bar{Z}$ in Porto’s notation [$ZX$ is short for $Y(ZX)\bar{Y}$], where $X$ and $Y$ denote linear polarization parallel and perpendicular to the crystallographic $a$ axis, respectively~[Fig.~\ref{Fig1_300K}(c)]. The $Z$ direction corresponds to the $c$-axis direction perpendicular to the (0~0~1) plane. 
Five scattering geometries were employed to probe excitations in different symmetry channels of HoAgGe. The relationship between the scattering geometries and the symmetry channels is given in Table~\ref{SymmetryAnalysis}.
         
\begin{table}[b]
\caption{\label{SymmetryAnalysis}The relationship between the scattering geometries and the symmetry channels for HoAgGe at zero magnetic field. $A'_1$, $A''_1$, $E'$ and $E''$ are the irreducible representations of the $D_{3h}$ point group. }
\begin{ruledtabular}
\begin{tabular}{cc}
Scattering geometry&Symmetry channel ($D_{3h}$)\\
\hline
$XX$&$A'_1+E'$\\
$XY$&$A''_1+E'$\\
$RR$&$A'_1+A''_1$\\
$RL$&$E'$\\
$ZX$&$E''$\\
\end{tabular}
\end{ruledtabular}
\end{table}

\begin{figure*}[!t] 
\begin{center}
\includegraphics[width=2\columnwidth]{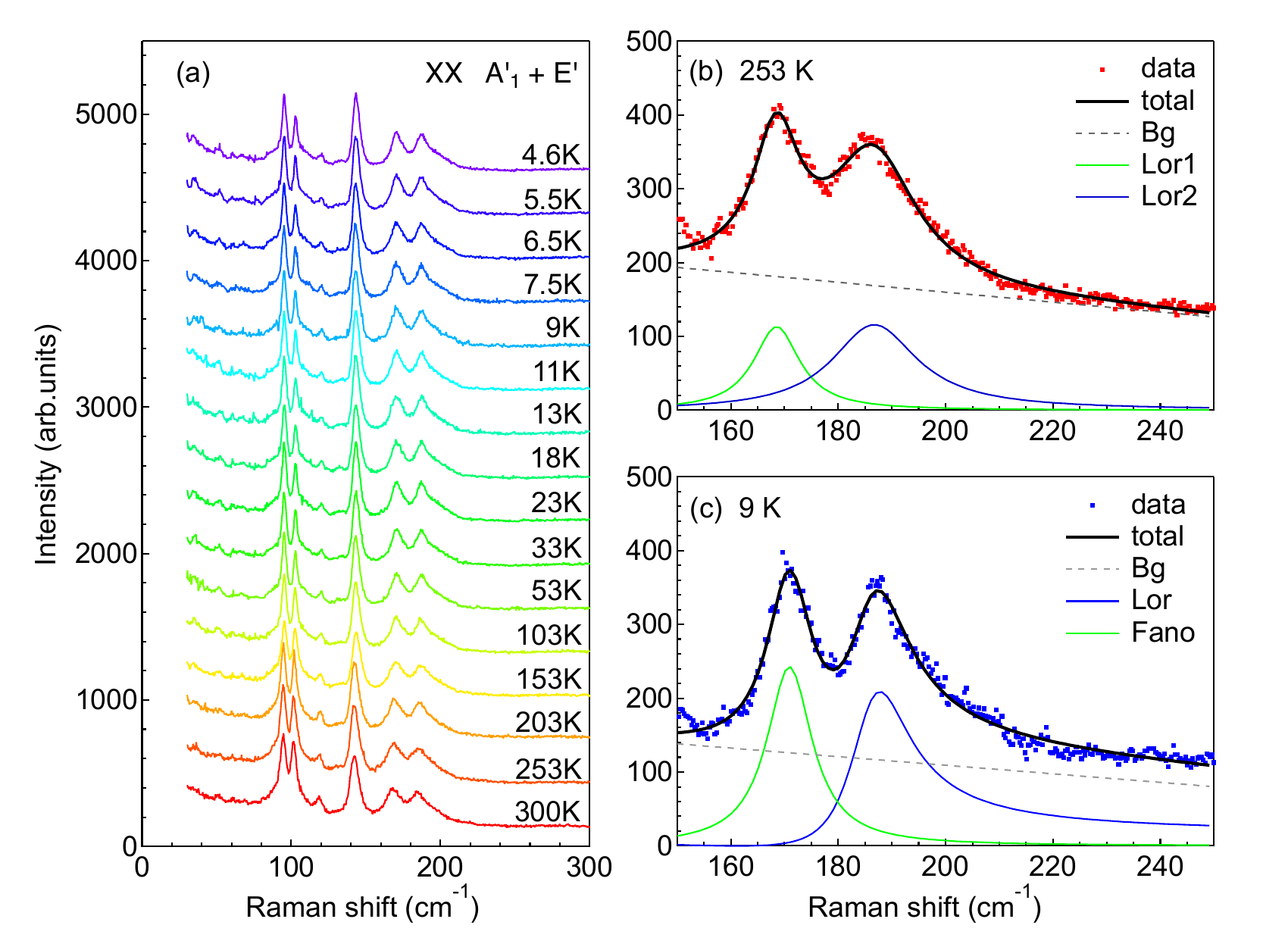}
\end{center}
\caption{\label{Fig2_T_dependence}
(a) Temperature dependence of the Raman response in the $XX$ ($A'_1+E'$) scattering geometry.
(b) Raman response for the two modes at 168 and 185\,\cm-1 at 300\,K. The solid black line is a fit using a sum of two Lorentzian functions and a linear background. The dashed line represents the background. The two components are shown by green and blue thin lines.
(c) Same as (b), but for 9\,K. The solid black line is a fit using a sum of a Lorentzian function, a Fano lineshape, and a linear background.
}
\end{figure*}

\textit{First-principles calculations.}\label{DFT} 
Density functional theory (DFT) calculations for HoAgGe were conducted using the QUANTUM ESPRESSO code~\cite{Giannozzi_2009_JPCM,Giannozzi_2017}. The exchange and correlation energies use the Perdew-Burke-Ernzerhof  form of the generalized gradient approximation~\cite{Perdew1996PhysRevLett}. The ground-state electronic structure was ascertained through the projector augmented wave method. A plane wave basis cutoff of 120\,Ry and charge density cutoff of 800\,Ry were employed, and the Monkhorst-Pack $k$-point grid was configured to be $8 \times 8 \times12$. The lattice structure was thoroughly relaxed until the total energy and forces acting on atoms dropped below 10$^{-8}$~Ry/atom and 10$^{-8}$~Ry/Bohr, respectively. The phonon dispersion of HoAgGe was calculated using scalar-relativistic density-functional perturbation theory with fully relativistic pseudopotentials from the PSEUDO-DOJO library~\cite{VANSETTEN201839}. 
PHONOPY~\cite{TOGO20151} was utilized to generate Raman dispersion, as well as to perform postprocessing on the calculated interatomic forces, thereby yielding the phonon frequencies.
The on-site Coulomb interaction $U$ is not included as it has a negligible effect on the Brillouin zone center phonon frequencies.

\textit{Group theoretical analysis.}\label{Group}
Group theoretical analyses were performed using the tool provided in the ISOTROPY software suite and the Bilbao Crystallographic Server \cite{Bilbao_1, Bilbao_4, Hatch2003}. The information about the irreducible representations of point groups and space groups follow the notations of Cracknell~$et~al$~\cite{cracknell1979general}.

\begin{table}[b]
\caption{\label{phonon_modes} 
Experimental phonon frequencies at the Brillouin zone center for HoAgGe at 300\,K and the phonon frequencies calculated by DFT. All the units are in \cm-1.}
\begin{ruledtabular}
\begin{tabular}{ccccc}
Symmetry &Activity&Maj. Displacement&Cal. &Exp.\\
\hline
$A_2''$&IR&Ho($z$), Ag($z$)	&86.3\\
$E''$&Raman&Ho($z$), Ag($z$)	&92.6&88\\
$E'$&IR+Raman&Ag($xy$), Ge($xy$)	&95.2&95\\
$A_1'$&Raman&Ho($xy$)	&95.8&-\\
$E'$&IR+Raman	&Ho($xy$), Ge($xy$)	&98.7&102\\
$E''$&Raman	&	Ho($z$)&106.9&104\\
$A_2'$&-	&Ho($xy$), Ag($xy$)	&107.2&\\
$E'$&IR+Raman&	Ho($xy$), Ag($xy$)	&123.2&119\\
$A_2'$&-	&Ho($xy$), Ag($xy$)		&146.4\\
$A_1'$&Raman	&Ag($xy$)&151.3&142\\
$A_1''$&-	&Ho($xy$), Ag($xy$)	&163.6\\
$E'$&IR+Raman&Ge($xy$)	&170.9&168\\
$A_2''$&IR&Ge($z$)		&180.5&\\
$A_2''$&IR&Ag($z$), Ge($z$)	&189.3\\
$E'$&IR+Raman&Ge($xy$)&194.9&185\\
$E'$&IR+Raman&Ge($xy$)	&206.3&-\\
\end{tabular}
\end{ruledtabular}
\end{table}  
										
\section{Results and Discussions}\label{Results} 
          
\subsection{Phonon modes}\label{Phonon}
                                                                                                                                                                                                                                                                                                                                                                                                         
HoAgGe belongs to the hexagonal structure with space group $P \bar{6}2m$ [No.~189, point group is $D_{3h}$, Fig.~\ref{Fig1_300K}(a)]. 
The Ho, Ag, Ge$^1$ (in the honeycomb layer), and Ge$^2$ atoms (in the kagome layer) have Wyckoff positions $3f$, $3d$, $2d$, and $1a$, respectively.                                                             
From the group theoretical considerations~\cite{Bilbao_1}, $\Gamma$-point phonon modes of the hexagonal HoAgGe can be expressed as $\Gamma_\text{total}$ = 2$A'_1$ $\oplus$ 2$A'_{2}$ $\oplus$ $A''_{1}$ $\oplus$ 4$A''_{2}$ $\oplus$ 7$E'$ $\oplus$ 2$E''$. Raman-active modes are $\Gamma_{\text{Raman}}$= 2$A'_1$ $\oplus$ 6$E'$ $\oplus$ 2$E''$, IR-active modes are $\Gamma_{\text{IR}}$=3$A''_{2}$ $\oplus$ 6$E'$, the acoustic mode is $\Gamma_{\text{acoustic}}$ =$A''_{2}$ $\oplus$ $E'$, and the silent modes are $\Gamma_{\text{silent}}$ = 2$A'_{2}$ $\oplus$ $A''_{1}$.
The following points should be noted:
(1) The inversion symmetry is absent in the point group $D_{3h}$. Thus, the $E'$ modes are both Raman and IR active, and $A''_{2}$ is IR active, while $A'_1$ is Raman active.
(2) Both Ag and Ho sites have $A'_1$ Raman-active modes while Ge sites have no $A'_1$ Raman-active modes.
(3) $E'$ modes are accessible from the $ab$-plane measurement while $E''$ modes are accessible from the $ac$-plane measurement.
                    
In Fig.~\ref{Fig1_300K}(e), we show the polarization-resolved Raman response for HoAgGe in five scattering geometries. In the $XX$ scattering geometry, six phonon modes (95, 102, 119, 142, 168, and 185\,\cm-1) are well resolved and are $A'_1$ and $E'$ symmetry modes. The symmetry of these modes can be separated by using the circular polarizations (Table~\ref{SymmetryAnalysis}). The single mode at 142\,\cm-1 observed in the $RR$ scattering geometry is assigned to be an $A'_1$ mode, which is mainly related to the Ag-lattice vibration modes. The remaining five modes (95, 102, 119, 168, and 185\,\cm-1) observed in the $RL$ scattering geometry are assigned to be $E'$ modes.  According to the group theoretical analysis, two $A'_1$ modes are expected. The one related to the Ho lattice vibration $A'_1$ mode is supposed to have a smaller vibration frequency due to the large mass of the Ho atom. It might be too weak to observe. In the $ZX$ scattering geometry, two modes observed at 88 and 104\,\cm-1 are assigned to be $E''$ modes. The phonon frequencies and symmetry of these modes are summarized in Table~\ref{phonon_modes}. 
{The major lattice vibration patterns for the Raman-active modes of HoAgGe are shown in Fig.~\ref{Fig2_pattern}.}
                                                                
To further understand the lattice dynamics of HoAgGe, we perform the DFT phonon calculations. We summarize the calculated phonon frequencies and the experimental  phonon frequencies at Brillouin zone center in Table~\ref{phonon_modes}. Overall, the experimentally observed phonon frequencies are consistent with the DFT phonon calculations. We note that two modes predicted by DFT are not detected in the experiments,  which might be due to the weak Raman intensities.

 \begin{figure*}[!t] 
\begin{center}
\includegraphics[width=2\columnwidth]{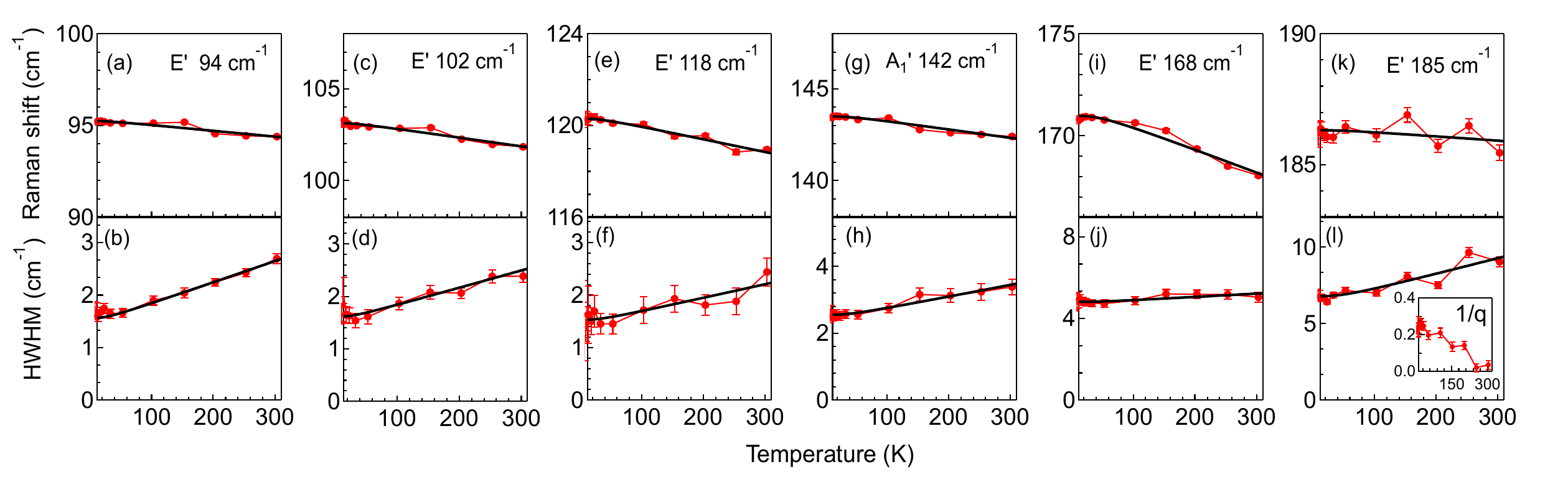}
\end{center}
\caption{\label{fitting_w_and_HMHM}
Fitting of the temperature dependence of the peak position, HWHM for the Raman modes of HoAgGe using the anharmonic phonon decay model above $T_2$ (see Appendix~\ref{Anharmonic_decay_model}) for the mode at (a) and (b) 94\,\cm-1, (c) and (d)  102\,\cm-1, (e) and (f) 118\,\cm-1, (g) and (h) 142\,\cm-1, (i) and (j) 168\,\cm-1, and (k) and (l) 185\,\cm-1. The inset of (l) shows the temperature dependence of the Fano model parameter $1/q$. The error bars represent one standard deviation. }
\end{figure*}

\begin{figure*}[!t] 
\begin{center}
\includegraphics[width=2\columnwidth]{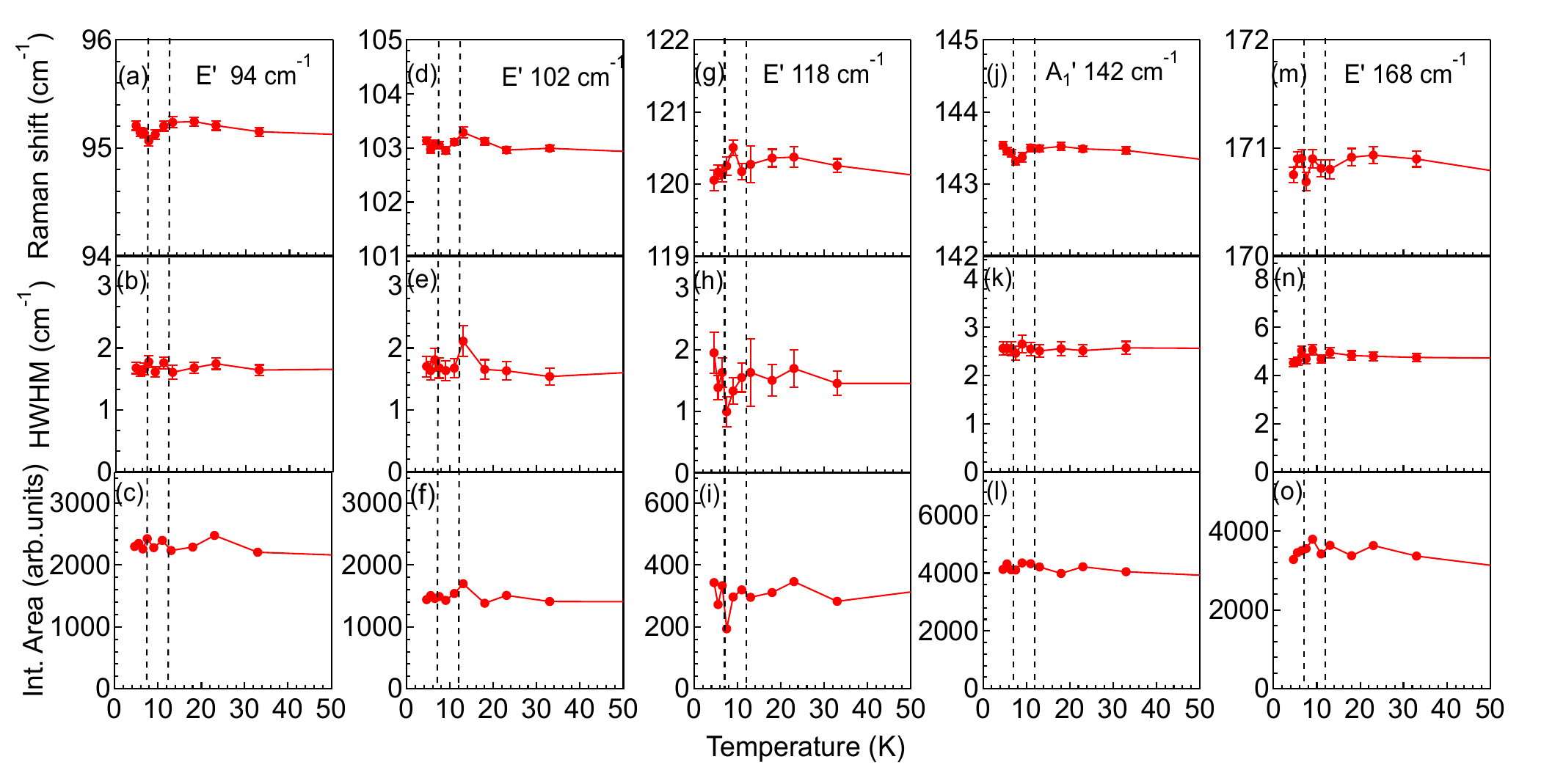}
\end{center}
\caption{\label{Fig3_fitting}
Temperature dependence of the peak position, HWHM, and integrated intensity for the Raman modes of HoAgGe for the mode at (a)-(c) 94\,\cm-1, (d)-(f) 102\,\cm-1,(g)-(i) 118\,\cm-1, (j)-(l) 142\,\cm-1, and (m)-(o) 168\,\cm-1. 
The error bars represent one standard deviation. The dashed lines mark temperatures $T_1$ and $T_2$.
}
\end{figure*}
                                                                                                                                                                                                                                                          
\subsection{Temperature dependence} \label{Temperature_dependence}                                                                      
                                                                                                                                                                                                                                               
After establishing the Raman modes of HoAgGe at room temperature, we present their temperature dependence.

In Fig.~\ref{Fig2_T_dependence}(a), we show the temperature dependence of the Raman response in the $XX$ ($A'_1$+$E'$) scattering geometry across the two magnetic phase transitions. All the phonon modes show hardening and narrowing upon cooling. 
{The temperature dependence of the mode frequency and half width at half maximum (HWHM) in the nonmagnetic phase shown in Fig.~\ref{fitting_w_and_HMHM} are consistent with the conventional phonon anharmonic decay model~[see Appendix~\ref{Anharmonic_decay_model}]}. Notably, as we show in Fig.~\ref{Fig2_T_dependence}(b), the $E'$ mode at 185\,\cm-1 shows a symmetric line shape at 253\,K, which can be modeled by a Lorentzian function. It develops an asymmetric line shape upon cooling, indicating Fano-type interference lineshape due to finite electron-phonon coupling. The Fano resonance effect originates from quantum interference between a discrete state (in this case a phonon mode) and an electron continuum, in which the excited eigenstates are a mixture of the discrete and continuum states~\cite{Fano_1961PhysRev}. 
The Fano lineshape is described by the equation                        
\begin{equation}
I(\omega)=A \frac{\left[q+\left(\omega-\omega_0\right) / \Gamma\right]^2}{1+\left[\left(\omega-\omega_0\right) / \Gamma\right]^2},
\end{equation}     
where $A$ represents the amplitude, $\omega_0$ represents the bare mode frequency,
$\Gamma$ is the HWHM of the coupled Fano-shaped mode, and $q$ is inversely proportional to the electron-phonon interaction strength.
As we show in Fig.~\ref{Fig2_T_dependence}(c), a Fano lineshape can model the mode at 185\,\cm-1 very well with parameters $\omega_0=185.9$, $\Gamma=6.7$, and $q=3.7$.
The strength of the electron-phonon coupling can be approximately estimated by the parameter $1/q$.  {The temperature dependence of $1/q$ is shown in the inset of Fig.~\ref{fitting_w_and_HMHM}(l). The parameter $1/q$ increases upon cooling}, and it is 0.27 at 9\,K for HoAgGe, similar to BaFe$_2$As$_2$ slightly below the structure phase transition ($|1/q|\sim0.35$)~\cite{Wu_2020PhysRevB}. 
The magnitude of $1/q$ in HoAgGe suggests moderate electron-phonon coupling in the Brillouin zone center.

In Fig.~\ref{Fig3_fitting}, we present the temperature dependence of the peak position, HWHM, and integrated intensity for the Raman modes of HoAgGe upon cooling across the magnetic phase transition.  The temperature dependence of the HWHM and the integrated intensity show small anomalies upon cooling across the magnetic phase transitions. In contrast, the temperature dependence of the phonon frequencies for the $E'$ mode at 94 and 102\,\cm-1 and the $A'_1$ mode at 142\,\cm-1 harden upon cooling to $T_2$, then softens below $T_2$, and finally harden again below $T_1$. The softening of both the $A'_1$ and $E'$ modes below $T_2$ is attributed to a renormalization of phonon energy due to spin-phonon coupling. Especially, the low-energy in-plane $E'$-symmetry mode containing mainly Ho lattice vibration modulates the neighboring Ho spin exchange energy, creating a coupling between the Ho lattice vibration mode and the Ho spin ordering. The magnitude of the phonon softening between $T_1$ and $T_2$ is about 0.2\,\cm-1, which is about $\delta \omega/\omega=-$0.2\%  for the mode at 102\,\cm-1.
It is similar in magnitude to the $E_g$ mode at 248\,\cm-1 in MnF$_2$ ($-0.5\%$)~\cite{Lockwood_1988JAP}, $E_g$ mode at 308\,\cm-1 in NiF$_2$ ($-0.3\%$)~\cite{Lockwood_2002LTP}, the 476 and the 573\,\cm-1 modes in Na$_{0.5}$CoO$_2$ ($-0.2\%$)~\cite{Zhang_2008PhysRevB}.
                                               
We can estimate the spin-phonon coupling constant by the shift of phonon frequency across the magnetic phase transitions via
\begin{equation}
\Delta \omega=\omega_p - \omega^0_p=
\lambda\left\langle\mathbf{S}_{\mathbf{i}} \cdot \mathbf{S}_{\mathbf{i}+1}\right\rangle.
\end{equation}
Here, $\left\langle \mathbf{S}_{\mathbf{i}} \cdot \mathbf{S}_{\mathbf{i}+1}\right\rangle$ denotes a statistical mechanical average for adjacent spin pairs. The spin-phonon coupling constant $\lambda$ is proportional to the second derivative of the direct exchange energy with respect to the phonon coordinate. It varies for each phonon and could have either a positive or negative sign depending on the phonon mode.
For a detailed comparison between experiment and theory,  the $T$-dependent  $\left\langle \mathbf{S}_{\mathbf{i}} \cdot \mathbf{S}_{\mathbf{i}+1}\right\rangle$ can be evaluated by $S^2 \phi(T)$ where $\phi(T)=|\left\langle \mathbf{S}_{\mathbf{i}} \cdot \mathbf{S}_{\mathbf{i}+1}\right\rangle|/S^2$. $S$ denotes the effective spin quantum number. 
$S^2$ can be estimated via the equation $S^2 = [M / (2 \mu_B)]^2$~\cite{Granado_1999PhysRevB,Zhang_2008PhysRevB,Hu_2019PhysRevB}, and $\phi(T)$ can be estimated from the $T$ dependence of the integrated intensities for the magnetic Bragg peak (1/3~1/3~0)~\cite{Zhao_2020_science}. 
At 4\,K, the ordered magnetic moment $M=7.5\mu_B$~\cite{Zhao_2020_science}, and the estimated $S^2$ is 14.
Since $\phi$(7\,K) $\sim 0.8$, and $\Delta \omega=-0.2 \pm 0.05$\,\cm-1 for the $E'$ mode at 102\,\cm-1,
the extracted spin-phonon coupling constant $\lambda \sim-0.02\pm0.004$\,\cm-1. This is similar for the modes at 94 and 142\,\cm-1. The overall spin-phonon coupling constant $\lambda$ in HoAgGe is about an order of magnitude smaller than those in the typical antiferromagnet NiF$_2$, FeF$_2$, and MnF$_2$, whose average $|\lambda|$ is close to 0.3-0.4~\cite{Lockwood_1988JAP,Lockwood_2002LTP}. 
It is much smaller than the ZnCr$_2$O$_4$ compound with a spin-phonon coupling constant of about 5\,\cm-1~\cite{Kant_2009PhysRevB}.
The small spin-phonon coupling constant in HoAgGe originates from a small phonon softening below the $T_2$ magnetic phase transition in a classical spin ice compound with a large ordered magnetic moment.

Finally, we discuss the implications of the finite spin-phonon coupling in HoAgGe. Zhao \textit{et al.} reported that the Hall conductivity $\sigma_{ac}$ change sign at metamagnetic transitions around 3\,T when magnetic fields are applied parallel to the kagome plane~\cite{Zhao_2024_Nature_Phy}. The sign change is believed to be related to the substantial changes in the band structure near the Fermi energy driven by the magnetic order change~\cite{Zhao_2024_Nature_Phy}. In the presence of finite spin-phonon coupling, the lattice structure in HoAgGe could respond to the magnetic order change, which might account for the band structure change in metamagnetic transitions. 
Furthermore, when an in-plane magnetic field is applied along the $b$ axis (e.g.~1.5\,T) for HoAgGe in the spin ice state below $T_1$, some of spins flip sign along the line between the corner and the center of the triangles. As a result, the magnetic unit cell changes from hexagonal at zero magnetic field to orthorhombic at finite magnetic field~\cite{Zhao_2020_science}. The lattice structure could respond to the new magnetic unit cell in the presence of the spin-lattice coupling, breaking the three-fold lattice rotational symmetry. A similar case was discussed for Tb$_2$Ti$_2$O$_7$~\cite{Jin_2020_PhysRevLett}.
Thus, the twofold degenerate $E'$ and $E''$ phonon modes in HoAgGe could split into two modes in the orthorhombic magnetic unit cell. 
Last, the magnetic field induced spin flips in HoAgGe also motivate us to study the potential emergent magnetic monopole and antimonopole pairs, as well as the attached electric dipoles in the presence of finite spin-phonon coupling.
While we do not observe such splitting of twofold degenerate modes and the signatures of the magnetic monopole and antimonopole pairs because our Raman experiment is performed at 0\,T, our results for the phonon modes at zero magnetic field and spin-phonon coupling provide a basis for the future study of the kagome spin-ice-related physics in the magnetic field.

\begin{figure}[!t] 
\begin{center}
\includegraphics[width=\columnwidth]{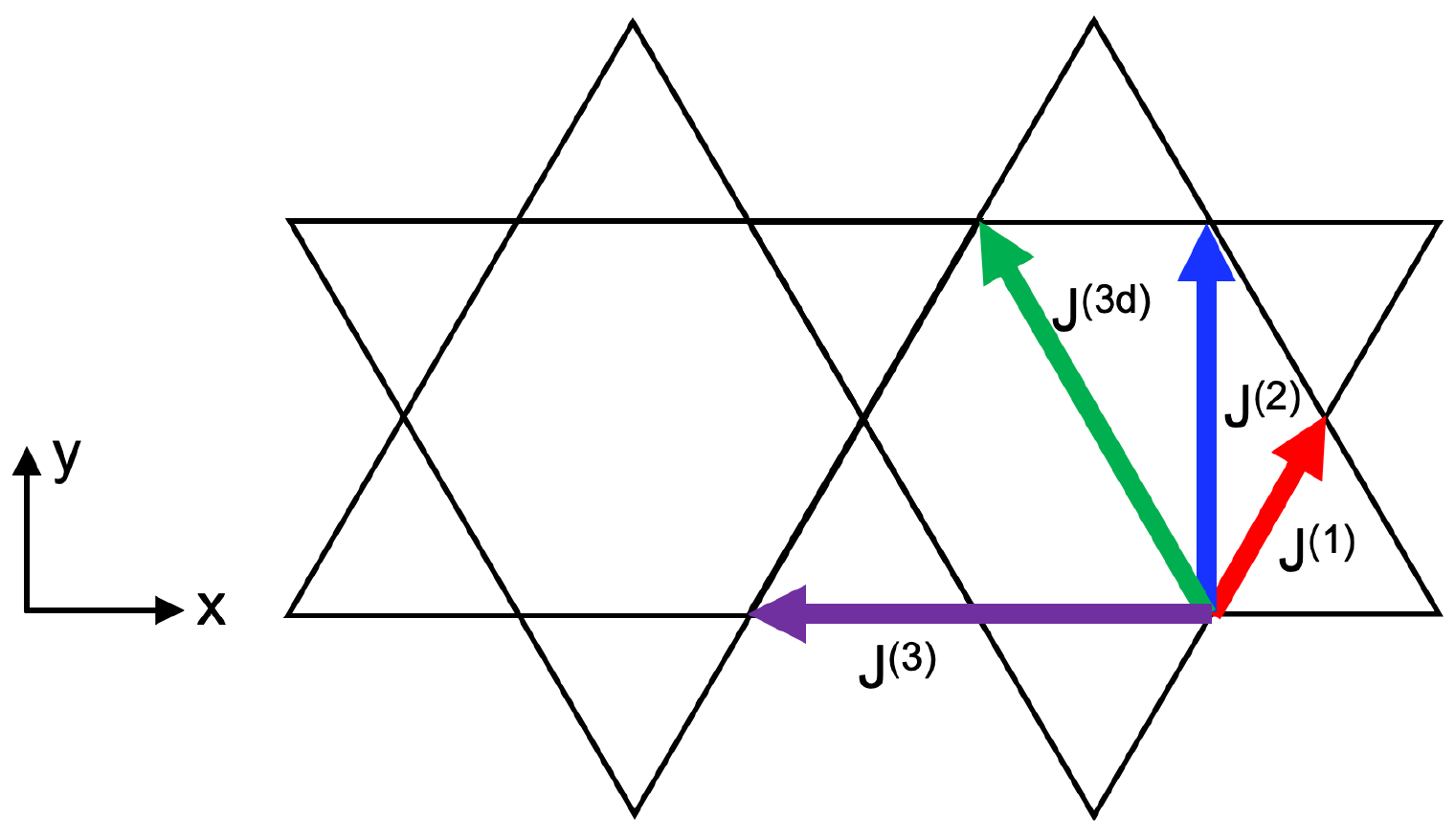}
\end{center}
\caption{\label{Jij}
Illustration of the kagome lattice. $J^{(1)}$, $J^{(2)}$, $J^{(3)}$, and $J^{(3d)}$ represent the nearest-neighbor exchange energies that are first, second, and third along the triangle edges and third along the hexagon diagonal, respectively.
}   
\end{figure}

{
Generally speaking, the existence of spin-phonon coupling in the spin ice system puts additional terms or constraints on the low-energy spin Hamiltonian. For HoAgGe, the low-energy spin Hamiltonian includes Heisenberg exchange couplings up to $n$th nearest neighbors, magnetic dipolar interaction, and Zeeman coupling with an external in-plane magnetic field:
\begin{equation}\label{Hamiltonian}
\begin{aligned}
H_0= & J_{0} \sum_{\langle i j\rangle^{(1)}} \vec{S}_i \cdot \vec{S}_j \\
& +D a^3 \sum_{i \neq j}\left[\frac{\vec{S}_i \cdot \vec{S}_j}{\left(r_{i j}^0\right)^3}-\frac{3\left(\vec{S}_i \cdot \hat{r}_{i j}\right)\left(\vec{S}_j \cdot \hat{r}_{i j}\right)}{\left(r_{i j}^0\right)^3}\right] \\
& -\vec{h} \cdot \sum_i \vec{S}_i,
\end{aligned}
\end{equation}
where $\vec{S}_i=S \sigma_i \hat{e}_i$ is a local Ising spin, $\sigma_i=\pm 1$ is the Ising variable, $\hat{e}_i$ is the local reference direction pointing to the center of the triangle, and $S$ is the spin magnitude.
$J_{0}$ is the strength of the first nearest-neighbor exchange coupling between site $i$ and $j$, and $D$ is the strength of the dipolar coupling. 
$r^0_{ij}$ is the distance between any pair of spins at site $i$ and site $j$.  $\hat{r}_{ij}$ is the unit vector from site $i$ to site $j$. $\vec{h}=h_x \hat{x}+h_y\hat{y}$ is the external magnetic field direction in the kagome plane~(Fig.~\ref{Jij}).
}

{
The effects of lattice deformations $\vec{u}_i$ on the magnetic order have been discussed in the literature for two models: the bond phonon model~\cite{Penc_2004_PhysRevLett} and Einstein site phonon model~\cite{Bergman_2006_PhysRevB,Wang_2008_PhysRevLett,Albarrac_2013_PhysRevB,Albarracin_2014JPC,Aoyama_2016_PhysRevLett,Pili_2019_PhysRevB,Gen_2022_PhysRevB,Gao_2024_PhysRevB,Ferrari_2021_PhysRevB,Ferrari_2024_PhysRevB}.
The bond phonon model assumes that each bond can deform independently which ignores the geometrical constraints~\cite{Penc_2004_PhysRevLett}. 
A more realistic phonon model can be described using the independent displacements of each atom at each site, with bond distances derived from these displacements, corresponding to the Einstein site phonon model~\cite{Bergman_2006_PhysRevB}.
If we adopt the standard Einstein site phonon model and perform a first-order Taylor expansion, the changes to the exchange and dipolar terms of the spin Hamiltonian as a result of atomic displacements can be approximated as linear terms in $\vec{u}_i$,
which can be integrated out to obtain an effective spin Hamiltonian.
Since the spin Hamiltonian Eq.~(\ref{Hamiltonian}) for HoAgGe is equivalent to the planar kagome spin ice model developed in the Ising pyrochlore system under an external magnetic field in the [111] direction, where the apical spins are completely aligned with the external magnetic field, we quote the effective spin Hamiltonian following Ref.~\cite{Albarrac_2013_PhysRevB}:
\begin{equation}
\begin{aligned}
\frac{H_{\mathrm{eff}}}{J_0 S^2}= & {\left[J_{\mathrm{eff}}^{(1)} \sum_{\langle i j\rangle^{(1)}} \sigma_i \sigma_j+J_{\mathrm{eff}}^{(2)} \sum_{\langle i j\rangle^{(2)}} \sigma_i \sigma_j\right.} \\
& \left.+J_{\mathrm{eff}}^{(3)} \sum_{\langle i j\rangle^{(3)}} \sigma_i \sigma_j+J_{\mathrm{eff}}^{(3 d)} \sum_{\langle i j\rangle^{(3 d)}} \sigma_i \sigma_j\right] \\
& -\sum_i \sigma_i\left[\tilde{h}_x\left(\hat{e}_i\right)_x+\tilde{h}_y\left(\hat{e}_i\right)_y\right]
\end{aligned}
\end{equation}
where $\langle i j\rangle^{(1)}$, $\langle i j\rangle^{(2)}$, $\langle i j\rangle^{(3)}$, and $\langle i j\rangle^{(3d)}$ are the 
first, second, and third neighbors along triangle edges and third neighbors along hexagon diagonal, respectively~(Fig.~\ref{Jij}). The spin-phonon coupling constant is defined as $\alpha=-\left.\frac{1}{J_0} \frac{\partial J}{\partial r_{i j}}\right|_{r_{i j}=r_{i j}^0}$. The dimensionless spin-phonon coupling constant can be defined as $\lambda'=a \alpha$ ($a$ is the distance between nearest neighbors).
The dimensionless dipolar interaction strength can be defined as $d=\frac{D}{J_0}$, and the dimensionless magnetic field can be defined as  $\tilde{h}_t =h_t /\left(J_0 S\right)$, with $t=x, y$. The effective exchange couplings are given by
 \begin{equation}\label{eqs}
\begin{aligned}
& J_{\mathrm{eff}}^{(1)}=-\frac{1}{2}+\frac{7}{4} d-\delta, \\
& J_{\mathrm{eff}}^{(2)}=-\frac{5}{12 \sqrt{3}} d+\delta, \\
& J_{\mathrm{eff}}^{(3 d)}=\frac{1}{8} d \\
& J_{\mathrm{eff}}^{(3)}=-\frac{5}{32} d+2 \delta,
\end{aligned}
\end{equation}
with $\delta =\frac{S^2}{4k}(-\frac{\lambda'}{2}+\frac{21}{4}d)^2 $~\cite{Albarrac_2013_PhysRevB}. From the effective exchange energy equations~\ref{eqs}, we find that the spin-phonon coupling affects both the first-, second-, and third-  neighbor exchange energies.
}

{
From the Monte Carlo simulations of HoAgGe~\cite{Zhao_2020_science},  $J^{(1)}$ is estimated to be $-$2\,meV, $J^{(2)}$ is estimated to be 0.23\,meV, and $J^{(3)}$ and $J^{(3d)}$ are estimated to be 0.127\,meV. The dipolar interaction has a typical energy scale of $\sim$0.11 meV. From our Raman results, the softening of the phonon modes at the magnetic ordering is about 0.2\,\cm-1, corresponding to an energy scale of about 0.025meV. It is about 1\% of $J^{(1)}$, 10\% of $J^{(2)}$,  20\% of $J^{(3)}$ and $J^{(3d)}$, and 20\% of the dipolar energy. While the spin-phonon coupling barely affects $J^{(1)}$ for HoAgGe, it moderately affects the further-nearest-neighbor exchange interactions. This may play a role in stabilizing the rich magnetic plateaus observed in HoAgGe~\cite{Zhao_2020_science} according to the kagome spin ice model~\cite{Albarrac_2013_PhysRevB}. More detailed theoretical studies of spin-phonon (spin-lattice) interaction by first-principles calculations~\cite{Mankovsky_2023_PhysRevB} or Monte Carlo simulations~\cite{Albarrac_2013_PhysRevB,Albarracin_2014JPC,Aoyama_2016_PhysRevLett,Pili_2019_PhysRevB,Gao_2024_PhysRevB} for the kagome spin ice HoAgGe on a quantitative level require future work.
}
\\
                                                                                                                                                                                                                                                                                                                                                                            
\section{Conclusions}\label{Conclusions} 
                             
In conclusion, we studied the lattice dynamics and spin-phonon coupling in the kagome spin ice compound HoAgGe using Raman spectroscopy and first-principles phonon calculations.
We detected 8 of 10 Raman-active phonon modes at 300\,K, the frequency of which is in accord with first-principles phonon calculations.
We found that upon colling, the low-energy phonon modes harden first but soften slightly below $T_2$ and, finally, harden again below $T_1$, suggesting finite spin-phonon coupling in HoAgGe. Furthermore, we found a high-energy $E'$ mode at 185\,\cm-1 that has an asymmetric line shape, indicating finite electron-phonon coupling in this system. The finite spin-phonon coupling and electron-phonon coupling motivate the study of the novel field-induced effects in HoAgGe, for example, spin-flips-induced threefold rotational symmetry breaking in the magnetic unit cell and emergent electric dipoles attached to the magnetic monopole and antimonopole pairs.

\begin{acknowledgments}
This work was supported by the National Natural Science Foundation of China (Grant No.~12404548 and No.~12274186), the National Key Research and Development Program of China (Grant No.~2022YFA1402704), the Strategic Priority Research Program of the Chinese Academy of Sciences (Grant No.~XDB33010100), and the Guangdong Provincial Quantum Science Strategic Initiative (No. GDZX2401011).      
The experimental work was supported by the Synergetic Extreme Condition User Facility (SECUF, https://cstr.cn/31123.02.SECUF).                                                                           
\end{acknowledgments}     
          
          
\appendix 

\section{Laser heating determination} \label{laser_heating_determination}                                                                                                                             

In the Raman experiments, the laser heating rate, a measure of the temperature increase per unit laser power (K/mW) in the focused laser spot, was determined by monitoring 
the phonon frequency anomalies induced by the magnetic order during the warm-up process with a constant laser power of 0.3\,mW. The temperature dependence of the phonon frequencies for HoAgGe is shown in Fig.~\ref{Fig3_fitting}. 
At a cryostat temperature of 8\,K, we detect a phonon anomaly corresponding to the partially magnetic ordered phase temperature $T_2$=11.6\,K, indicating the laser spot temperature is close to $T_2$. 
Thus, the heating coefficient can be determined via $8\,\text{K}+0.3\,\text{mW} \times \text{k} \approx 11.6$\,K. In this way, we deduce the heating coefficient $k \approx 12\pm1$\,K/mW.

\section{Anharmonic phonon decay model}\label{Anharmonic_decay_model}    

{
In this appendix, we discuss the anharmonic phonon decay model.
We fit the temperature dependence of the phonon frequency and HWHM by the anharmonic phonon decay model~\cite{Klemens_PhysRev148,Cardona_PRB1984}: 
\begin{equation}
\label{eq_omega1}
\omega_1(T)=\omega_{0}-C_1\{1+ 2n[\Omega(T)/2]\},\\
\end{equation}
\begin{equation}
\label{eq_gamma1}
\Gamma_1(T)=\gamma_{0}+\gamma_1\{1+ 2n[\Omega(T)/2]\},
\end{equation}
where $\Omega(T)= \hbar \omega / k_BT$ and $n(x)=1/(e^x-1)$ is the Bose-Einstein distribution function. $\omega_1(T)$ and $\Gamma_1(T)$ involve mainly a three-phonon decay process in which an optical phonon decays into two acoustic modes with equal energies and opposite momenta. 
}
      

\begin{thebibliography}{54}%
\makeatletter
\providecommand \@ifxundefined [1]{%
 \@ifx{#1\undefined}
}%
\providecommand \@ifnum [1]{%
 \ifnum #1\expandafter \@firstoftwo
 \else \expandafter \@secondoftwo
 \fi
}%
\providecommand \@ifx [1]{%
 \ifx #1\expandafter \@firstoftwo
 \else \expandafter \@secondoftwo
 \fi
}%
\providecommand \natexlab [1]{#1}%
\providecommand \enquote  [1]{``#1''}%
\providecommand \bibnamefont  [1]{#1}%
\providecommand \bibfnamefont [1]{#1}%
\providecommand \citenamefont [1]{#1}%
\providecommand \href@noop [0]{\@secondoftwo}%
\providecommand \href [0]{\begingroup \@sanitize@url \@href}%
\providecommand \@href[1]{\@@startlink{#1}\@@href}%
\providecommand \@@href[1]{\endgroup#1\@@endlink}%
\providecommand \@sanitize@url [0]{\catcode `\\12\catcode `\$12\catcode
  `\&12\catcode `\#12\catcode `\^12\catcode `\_12\catcode `\%12\relax}%
\providecommand \@@startlink[1]{}%
\providecommand \@@endlink[0]{}%
\providecommand \url  [0]{\begingroup\@sanitize@url \@url }%
\providecommand \@url [1]{\endgroup\@href {#1}{\urlprefix }}%
\providecommand \urlprefix  [0]{URL }%
\providecommand \Eprint [0]{\href }%
\providecommand \doibase [0]{http://dx.doi.org/}%
\providecommand \selectlanguage [0]{\@gobble}%
\providecommand \bibinfo  [0]{\@secondoftwo}%
\providecommand \bibfield  [0]{\@secondoftwo}%
\providecommand \translation [1]{[#1]}%
\providecommand \BibitemOpen [0]{}%
\providecommand \bibitemStop [0]{}%
\providecommand \bibitemNoStop [0]{.\EOS\space}%
\providecommand \EOS [0]{\spacefactor3000\relax}%
\providecommand \BibitemShut  [1]{\csname bibitem#1\endcsname}%
\let\auto@bib@innerbib\@empty
\bibitem [{\citenamefont {Sy\^ozi}(1951)}]{Itiro1951}%
  \BibitemOpen
  \bibfield  {author} {\bibinfo {author} {\bibfnamefont {Itiro}\ \bibnamefont
  {Sy\^ozi}},\ }\bibfield  {title} {\enquote {\bibinfo {title} {{Statistics of
  Kagome Lattice}},}\ }\href {\doibase 10.1143/ptp/6.3.306} {\bibfield
  {journal} {\bibinfo  {journal} {Progress of Theoretical Physics}\ }\textbf
  {\bibinfo {volume} {6}},\ \bibinfo {pages} {306--308} (\bibinfo {year}
  {1951})}\BibitemShut {NoStop}%
\bibitem [{\citenamefont {Gilbert}\ \emph {et~al.}(2016)\citenamefont
  {Gilbert}, \citenamefont {Nisoli},\ and\ \citenamefont
  {Schiffer}}]{Gilbert_2016_PhysicsToday}%
  \BibitemOpen
  \bibfield  {author} {\bibinfo {author} {\bibfnamefont {Ian}\ \bibnamefont
  {Gilbert}}, \bibinfo {author} {\bibfnamefont {Cristiano}\ \bibnamefont
  {Nisoli}}, \ and\ \bibinfo {author} {\bibfnamefont {Peter}\ \bibnamefont
  {Schiffer}},\ }\bibfield  {title} {\enquote {\bibinfo {title} {{Frustration
  by design}},}\ }\href {\doibase 10.1063/PT.3.3237} {\bibfield  {journal}
  {\bibinfo  {journal} {Physics Today}\ }\textbf {\bibinfo {volume} {69}},\
  \bibinfo {pages} {54--59} (\bibinfo {year} {2016})}\BibitemShut {NoStop}%
\bibitem [{\citenamefont {Balents}(2010)}]{Balents2010}%
  \BibitemOpen
  \bibfield  {author} {\bibinfo {author} {\bibfnamefont {Leon}\ \bibnamefont
  {Balents}},\ }\bibfield  {title} {\enquote {\bibinfo {title} {Spin liquids in
  frustrated magnets},}\ }\href {\doibase 10.1038/nature08917} {\bibfield
  {journal} {\bibinfo  {journal} {Nature}\ }\textbf {\bibinfo {volume} {464}},\
  \bibinfo {pages} {199--208} (\bibinfo {year} {2010})}\BibitemShut {NoStop}%
\bibitem [{\citenamefont {Broholm}\ \emph {et~al.}(2020)\citenamefont
  {Broholm}, \citenamefont {Cava}, \citenamefont {Kivelson}, \citenamefont
  {Nocera}, \citenamefont {Norman},\ and\ \citenamefont
  {Senthil}}]{Broholm_2020_Science}%
  \BibitemOpen
  \bibfield  {author} {\bibinfo {author} {\bibfnamefont {C.}~\bibnamefont
  {Broholm}}, \bibinfo {author} {\bibfnamefont {R.~J.}\ \bibnamefont {Cava}},
  \bibinfo {author} {\bibfnamefont {S.~A.}\ \bibnamefont {Kivelson}}, \bibinfo
  {author} {\bibfnamefont {D.~G.}\ \bibnamefont {Nocera}}, \bibinfo {author}
  {\bibfnamefont {M.~R.}\ \bibnamefont {Norman}}, \ and\ \bibinfo {author}
  {\bibfnamefont {T.}~\bibnamefont {Senthil}},\ }\bibfield  {title} {\enquote
  {\bibinfo {title} {Quantum spin liquids},}\ }\href {\doibase
  10.1126/science.aay0668} {\bibfield  {journal} {\bibinfo  {journal}
  {Science}\ }\textbf {\bibinfo {volume} {367}},\ \bibinfo {pages} {eaay0668}
  (\bibinfo {year} {2020})}\BibitemShut {NoStop}%
\bibitem [{\citenamefont {Bramwell}\ and\ \citenamefont
  {Gingras}(2001)}]{Steven_2001_Science}%
  \BibitemOpen
  \bibfield  {author} {\bibinfo {author} {\bibfnamefont {Steven~T.}\
  \bibnamefont {Bramwell}}\ and\ \bibinfo {author} {\bibfnamefont {Michel
  J.~P.}\ \bibnamefont {Gingras}},\ }\bibfield  {title} {\enquote {\bibinfo
  {title} {Spin ice state in frustrated magnetic pyrochlore materials},}\
  }\href {\doibase 10.1126/science.1064761} {\bibfield  {journal} {\bibinfo
  {journal} {Science}\ }\textbf {\bibinfo {volume} {294}},\ \bibinfo {pages}
  {1495--1501} (\bibinfo {year} {2001})}\BibitemShut {NoStop}%
\bibitem [{\citenamefont {Castelnovo}\ \emph {et~al.}(2012)\citenamefont
  {Castelnovo}, \citenamefont {Moessner},\ and\ \citenamefont
  {Sondhi}}]{Castelnovo_2012_review}%
  \BibitemOpen
  \bibfield  {author} {\bibinfo {author} {\bibfnamefont {C.}~\bibnamefont
  {Castelnovo}}, \bibinfo {author} {\bibfnamefont {R.}~\bibnamefont
  {Moessner}}, \ and\ \bibinfo {author} {\bibfnamefont {S.L.}\ \bibnamefont
  {Sondhi}},\ }\bibfield  {title} {\enquote {\bibinfo {title} {Spin ice,
  fractionalization, and topological order},}\ }\href {\doibase
  https://doi.org/10.1146/annurev-conmatphys-020911-125058} {\bibfield
  {journal} {\bibinfo  {journal} {Annual Review of Condensed Matter Physics}\
  }\textbf {\bibinfo {volume} {3}},\ \bibinfo {pages} {35--55} (\bibinfo {year}
  {2012})}\BibitemShut {NoStop}%
\bibitem [{\citenamefont {Castelnovo}\ \emph {et~al.}(2008)\citenamefont
  {Castelnovo}, \citenamefont {Moessner},\ and\ \citenamefont
  {Sondhi}}]{Castelnovo2008}%
  \BibitemOpen
  \bibfield  {author} {\bibinfo {author} {\bibfnamefont {C.}~\bibnamefont
  {Castelnovo}}, \bibinfo {author} {\bibfnamefont {R.}~\bibnamefont
  {Moessner}}, \ and\ \bibinfo {author} {\bibfnamefont {S.~L.}\ \bibnamefont
  {Sondhi}},\ }\bibfield  {title} {\enquote {\bibinfo {title} {Magnetic
  monopoles in spin ice},}\ }\href {\doibase 10.1038/nature06433} {\bibfield
  {journal} {\bibinfo  {journal} {Nature}\ }\textbf {\bibinfo {volume} {451}},\
  \bibinfo {pages} {42--45} (\bibinfo {year} {2008})}\BibitemShut {NoStop}%
\bibitem [{\citenamefont {Bramwell}\ and\ \citenamefont
  {Harris}(2020)}]{Bramwell_2020}%
  \BibitemOpen
  \bibfield  {author} {\bibinfo {author} {\bibfnamefont {Steven~T.}\
  \bibnamefont {Bramwell}}\ and\ \bibinfo {author} {\bibfnamefont {Mark~J.}\
  \bibnamefont {Harris}},\ }\bibfield  {title} {\enquote {\bibinfo {title} {The
  history of spin ice},}\ }\href {\doibase 10.1088/1361-648X/ab8423} {\bibfield
   {journal} {\bibinfo  {journal} {Journal of Physics: Condensed Matter}\
  }\textbf {\bibinfo {volume} {32}},\ \bibinfo {pages} {374010} (\bibinfo
  {year} {2020})}\BibitemShut {NoStop}%
\bibitem [{\citenamefont {Khomskii}(2010)}]{Khomskii_2010}%
  \BibitemOpen
  \bibfield  {author} {\bibinfo {author} {\bibfnamefont {D.~I.}\ \bibnamefont
  {Khomskii}},\ }\bibfield  {title} {\enquote {\bibinfo {title} {{Spin
  chirality and nontrivial charge dynamics in frustrated Mott insulators:
  spontaneous currents and charge redistribution}},}\ }\href {\doibase
  10.1088/0953-8984/22/16/164209} {\bibfield  {journal} {\bibinfo  {journal}
  {Journal of Physics: Condensed Matter}\ }\textbf {\bibinfo {volume} {22}},\
  \bibinfo {pages} {164209} (\bibinfo {year} {2010})}\BibitemShut {NoStop}%
\bibitem [{\citenamefont {Khomskii}(2012)}]{Khomskii2012}%
  \BibitemOpen
  \bibfield  {author} {\bibinfo {author} {\bibfnamefont {D.~I.}\ \bibnamefont
  {Khomskii}},\ }\bibfield  {title} {\enquote {\bibinfo {title} {Electric
  dipoles on magnetic monopoles in spin ice},}\ }\href {\doibase
  10.1038/ncomms1904} {\bibfield  {journal} {\bibinfo  {journal} {Nature
  Communications}\ }\textbf {\bibinfo {volume} {3}},\ \bibinfo {pages} {904}
  (\bibinfo {year} {2012})}\BibitemShut {NoStop}%
\bibitem [{\citenamefont {Grams}\ \emph {et~al.}(2014)\citenamefont {Grams},
  \citenamefont {Valldor}, \citenamefont {Garst},\ and\ \citenamefont
  {Hemberger}}]{Grams_2014_NC}%
  \BibitemOpen
  \bibfield  {author} {\bibinfo {author} {\bibfnamefont {Christoph~P.}\
  \bibnamefont {Grams}}, \bibinfo {author} {\bibfnamefont {Martin}\
  \bibnamefont {Valldor}}, \bibinfo {author} {\bibfnamefont {Markus}\
  \bibnamefont {Garst}}, \ and\ \bibinfo {author} {\bibfnamefont {Joachim}\
  \bibnamefont {Hemberger}},\ }\bibfield  {title} {\enquote {\bibinfo {title}
  {Critical speeding-up in the magnetoelectric response of spin-ice near its
  monopole liquid--gas transition},}\ }\href {\doibase 10.1038/ncomms5853}
  {\bibfield  {journal} {\bibinfo  {journal} {Nature Communications}\ }\textbf
  {\bibinfo {volume} {5}},\ \bibinfo {pages} {4853} (\bibinfo {year}
  {2014})}\BibitemShut {NoStop}%
\bibitem [{\citenamefont {Aleksandrov}\ \emph {et~al.}(1985)\citenamefont
  {Aleksandrov}, \citenamefont {Lidskii}, \citenamefont {Mamsurova},
  \citenamefont {Neigauz}, \citenamefont {Pigal'skii}, \citenamefont {Pukhov},
  \citenamefont {Trusevich},\ and\ \citenamefont
  {Shcherbakova}}]{aleksandrov1985crystal}%
  \BibitemOpen
  \bibfield  {author} {\bibinfo {author} {\bibfnamefont {I.~V.}\ \bibnamefont
  {Aleksandrov}}, \bibinfo {author} {\bibfnamefont {B.~V.}\ \bibnamefont
  {Lidskii}}, \bibinfo {author} {\bibfnamefont {L.~G.}\ \bibnamefont
  {Mamsurova}}, \bibinfo {author} {\bibfnamefont {M.~G.}\ \bibnamefont
  {Neigauz}}, \bibinfo {author} {\bibfnamefont {K.~S.}\ \bibnamefont
  {Pigal'skii}}, \bibinfo {author} {\bibfnamefont {K.~K.}\ \bibnamefont
  {Pukhov}}, \bibinfo {author} {\bibfnamefont {N.~G.}\ \bibnamefont
  {Trusevich}}, \ and\ \bibinfo {author} {\bibfnamefont {L.~G.}\ \bibnamefont
  {Shcherbakova}},\ }\bibfield  {title} {\enquote {\bibinfo {title} {Crystal
  field effects and the nature of the giant magnetostriction in terbium
  dititanate},}\ }\href@noop {} {\bibfield  {journal} {\bibinfo  {journal}
  {Soviet Journal of Experimental and Theoretical Physics}\ }\textbf {\bibinfo
  {volume} {62}},\ \bibinfo {pages} {1287} (\bibinfo {year}
  {1985})}\BibitemShut {NoStop}%
\bibitem [{\citenamefont {Ruff}\ \emph {et~al.}(2010)\citenamefont {Ruff},
  \citenamefont {Islam}, \citenamefont {Clancy}, \citenamefont {Ross},
  \citenamefont {Nojiri}, \citenamefont {Matsuda}, \citenamefont {Dabkowska},
  \citenamefont {Dabkowski},\ and\ \citenamefont
  {Gaulin}}]{Ruff_2010_PhysRevLett}%
  \BibitemOpen
  \bibfield  {author} {\bibinfo {author} {\bibfnamefont {J.~P.~C.}\
  \bibnamefont {Ruff}}, \bibinfo {author} {\bibfnamefont {Z.}~\bibnamefont
  {Islam}}, \bibinfo {author} {\bibfnamefont {J.~P.}\ \bibnamefont {Clancy}},
  \bibinfo {author} {\bibfnamefont {K.~A.}\ \bibnamefont {Ross}}, \bibinfo
  {author} {\bibfnamefont {H.}~\bibnamefont {Nojiri}}, \bibinfo {author}
  {\bibfnamefont {Y.~H.}\ \bibnamefont {Matsuda}}, \bibinfo {author}
  {\bibfnamefont {H.~A.}\ \bibnamefont {Dabkowska}}, \bibinfo {author}
  {\bibfnamefont {A.~D.}\ \bibnamefont {Dabkowski}}, \ and\ \bibinfo {author}
  {\bibfnamefont {B.~D.}\ \bibnamefont {Gaulin}},\ }\bibfield  {title}
  {\enquote {\bibinfo {title} {{Magnetoelastics of a Spin Liquid: X-Ray
  Diffraction Studies of ${\mathrm{Tb}}_{2}{\mathrm{Ti}}_{2}{\mathbf{O}}_{7}$
  in Pulsed Magnetic Fields}},}\ }\href {\doibase
  10.1103/PhysRevLett.105.077203} {\bibfield  {journal} {\bibinfo  {journal}
  {Phys. Rev. Lett.}\ }\textbf {\bibinfo {volume} {105}},\ \bibinfo {pages}
  {077203} (\bibinfo {year} {2010})}\BibitemShut {NoStop}%
\bibitem [{\citenamefont {Klekovkina}\ \emph {et~al.}(2011)\citenamefont
  {Klekovkina}, \citenamefont {Zakirov}, \citenamefont {Malkin},\ and\
  \citenamefont {Kasatkina}}]{Klekovkina_2011}%
  \BibitemOpen
  \bibfield  {author} {\bibinfo {author} {\bibfnamefont {V.~V.}\ \bibnamefont
  {Klekovkina}}, \bibinfo {author} {\bibfnamefont {A.~R.}\ \bibnamefont
  {Zakirov}}, \bibinfo {author} {\bibfnamefont {B.~Z.}\ \bibnamefont {Malkin}},
  \ and\ \bibinfo {author} {\bibfnamefont {L.~A.}\ \bibnamefont {Kasatkina}},\
  }\bibfield  {title} {\enquote {\bibinfo {title} {{Simulations of magnetic and
  magnetoelastic properties of Tb$_2$Ti$_2$O$_7$ in paramagnetic phase}},}\
  }\href {\doibase 10.1088/1742-6596/324/1/012036} {\bibfield  {journal}
  {\bibinfo  {journal} {Journal of Physics: Conference Series}\ }\textbf
  {\bibinfo {volume} {324}},\ \bibinfo {pages} {012036} (\bibinfo {year}
  {2011})}\BibitemShut {NoStop}%
\bibitem [{\citenamefont {Fennell}\ \emph {et~al.}(2014)\citenamefont
  {Fennell}, \citenamefont {Kenzelmann}, \citenamefont {Roessli}, \citenamefont
  {Mutka}, \citenamefont {Ollivier}, \citenamefont {Ruminy}, \citenamefont
  {Stuhr}, \citenamefont {Zaharko}, \citenamefont {Bovo}, \citenamefont
  {Cervellino}, \citenamefont {Haas},\ and\ \citenamefont
  {Cava}}]{Fennell_2014_PhysRevLett}%
  \BibitemOpen
  \bibfield  {author} {\bibinfo {author} {\bibfnamefont {T.}~\bibnamefont
  {Fennell}}, \bibinfo {author} {\bibfnamefont {M.}~\bibnamefont {Kenzelmann}},
  \bibinfo {author} {\bibfnamefont {B.}~\bibnamefont {Roessli}}, \bibinfo
  {author} {\bibfnamefont {H.}~\bibnamefont {Mutka}}, \bibinfo {author}
  {\bibfnamefont {J.}~\bibnamefont {Ollivier}}, \bibinfo {author}
  {\bibfnamefont {M.}~\bibnamefont {Ruminy}}, \bibinfo {author} {\bibfnamefont
  {U.}~\bibnamefont {Stuhr}}, \bibinfo {author} {\bibfnamefont
  {O.}~\bibnamefont {Zaharko}}, \bibinfo {author} {\bibfnamefont
  {L.}~\bibnamefont {Bovo}}, \bibinfo {author} {\bibfnamefont {A.}~\bibnamefont
  {Cervellino}}, \bibinfo {author} {\bibfnamefont {M.~K.}\ \bibnamefont
  {Haas}}, \ and\ \bibinfo {author} {\bibfnamefont {R.~J.}\ \bibnamefont
  {Cava}},\ }\bibfield  {title} {\enquote {\bibinfo {title} {{Magnetoelastic
  Excitations in the Pyrochlore Spin Liquid
  ${\mathrm{Tb}}_{2}{\mathrm{Ti}}_{2}{\mathbf{O}}_{7}$}},}\ }\href {\doibase
  10.1103/PhysRevLett.112.017203} {\bibfield  {journal} {\bibinfo  {journal}
  {Phys. Rev. Lett.}\ }\textbf {\bibinfo {volume} {112}},\ \bibinfo {pages}
  {017203} (\bibinfo {year} {2014})}\BibitemShut {NoStop}%
\bibitem [{\citenamefont {Ruminy}\ \emph {et~al.}(2019)\citenamefont {Ruminy},
  \citenamefont {Guitteny}, \citenamefont {Robert}, \citenamefont {Regnault},
  \citenamefont {Boehm}, \citenamefont {Steffens}, \citenamefont {Mutka},
  \citenamefont {Ollivier}, \citenamefont {Stuhr}, \citenamefont {White},
  \citenamefont {Roessli}, \citenamefont {Bovo}, \citenamefont {Decorse},
  \citenamefont {Haas}, \citenamefont {Cava}, \citenamefont {Mirebeau},
  \citenamefont {Kenzelmann}, \citenamefont {Petit},\ and\ \citenamefont
  {Fennell}}]{Ruminy_2019_PhysRevB}%
  \BibitemOpen
  \bibfield  {author} {\bibinfo {author} {\bibfnamefont {M.}~\bibnamefont
  {Ruminy}}, \bibinfo {author} {\bibfnamefont {S.}~\bibnamefont {Guitteny}},
  \bibinfo {author} {\bibfnamefont {J.}~\bibnamefont {Robert}}, \bibinfo
  {author} {\bibfnamefont {L.-P.}\ \bibnamefont {Regnault}}, \bibinfo {author}
  {\bibfnamefont {M.}~\bibnamefont {Boehm}}, \bibinfo {author} {\bibfnamefont
  {P.}~\bibnamefont {Steffens}}, \bibinfo {author} {\bibfnamefont
  {H.}~\bibnamefont {Mutka}}, \bibinfo {author} {\bibfnamefont
  {J.}~\bibnamefont {Ollivier}}, \bibinfo {author} {\bibfnamefont
  {U.}~\bibnamefont {Stuhr}}, \bibinfo {author} {\bibfnamefont {J.~S.}\
  \bibnamefont {White}}, \bibinfo {author} {\bibfnamefont {B.}~\bibnamefont
  {Roessli}}, \bibinfo {author} {\bibfnamefont {L.}~\bibnamefont {Bovo}},
  \bibinfo {author} {\bibfnamefont {C.}~\bibnamefont {Decorse}}, \bibinfo
  {author} {\bibfnamefont {M.~K.}\ \bibnamefont {Haas}}, \bibinfo {author}
  {\bibfnamefont {R.~J.}\ \bibnamefont {Cava}}, \bibinfo {author}
  {\bibfnamefont {I.}~\bibnamefont {Mirebeau}}, \bibinfo {author}
  {\bibfnamefont {M.}~\bibnamefont {Kenzelmann}}, \bibinfo {author}
  {\bibfnamefont {S.}~\bibnamefont {Petit}}, \ and\ \bibinfo {author}
  {\bibfnamefont {T.}~\bibnamefont {Fennell}},\ }\bibfield  {title} {\enquote
  {\bibinfo {title} {{Magnetoelastic excitation spectrum in the rare-earth
  pyrochlore ${\mathrm{Tb}}_{2}{\mathrm{Ti}}_{2}{\mathrm{O}}_{7}$}},}\ }\href
  {\doibase 10.1103/PhysRevB.99.224431} {\bibfield  {journal} {\bibinfo
  {journal} {Phys. Rev. B}\ }\textbf {\bibinfo {volume} {99}},\ \bibinfo
  {pages} {224431} (\bibinfo {year} {2019})}\BibitemShut {NoStop}%
\bibitem [{\citenamefont {Jin}\ \emph {et~al.}(2020)\citenamefont {Jin},
  \citenamefont {Liu}, \citenamefont {Chang}, \citenamefont {Zhang},
  \citenamefont {Wang}, \citenamefont {Liu}, \citenamefont {Wang},
  \citenamefont {Sun}, \citenamefont {Chen}, \citenamefont {Sun},\ and\
  \citenamefont {Zhang}}]{Jin_2020_PhysRevLett}%
  \BibitemOpen
  \bibfield  {author} {\bibinfo {author} {\bibfnamefont {Feng}\ \bibnamefont
  {Jin}}, \bibinfo {author} {\bibfnamefont {Changle}\ \bibnamefont {Liu}},
  \bibinfo {author} {\bibfnamefont {Yanfen}\ \bibnamefont {Chang}}, \bibinfo
  {author} {\bibfnamefont {Anmin}\ \bibnamefont {Zhang}}, \bibinfo {author}
  {\bibfnamefont {Yimeng}\ \bibnamefont {Wang}}, \bibinfo {author}
  {\bibfnamefont {Weiwei}\ \bibnamefont {Liu}}, \bibinfo {author}
  {\bibfnamefont {Xiaoqun}\ \bibnamefont {Wang}}, \bibinfo {author}
  {\bibfnamefont {Young}\ \bibnamefont {Sun}}, \bibinfo {author} {\bibfnamefont
  {Gang}\ \bibnamefont {Chen}}, \bibinfo {author} {\bibfnamefont {Xuefeng}\
  \bibnamefont {Sun}}, \ and\ \bibinfo {author} {\bibfnamefont {Qingming}\
  \bibnamefont {Zhang}},\ }\bibfield  {title} {\enquote {\bibinfo {title}
  {{Experimental Identification of Electric Dipoles Induced by Magnetic
  Monopoles in ${\mathrm{Tb}}_{2}{\mathrm{Ti}}_{2}{\mathrm{O}}_{7}$}},}\ }\href
  {\doibase 10.1103/PhysRevLett.124.087601} {\bibfield  {journal} {\bibinfo
  {journal} {Phys. Rev. Lett.}\ }\textbf {\bibinfo {volume} {124}},\ \bibinfo
  {pages} {087601} (\bibinfo {year} {2020})}\BibitemShut {NoStop}%
\bibitem [{\citenamefont {Zhao}\ \emph {et~al.}(2020)\citenamefont {Zhao},
  \citenamefont {Deng}, \citenamefont {Chen}, \citenamefont {Ross},
  \citenamefont {Petříček}, \citenamefont {Günther}, \citenamefont
  {Russina}, \citenamefont {Hutanu},\ and\ \citenamefont
  {Gegenwart}}]{Zhao_2020_science}%
  \BibitemOpen
  \bibfield  {author} {\bibinfo {author} {\bibfnamefont {Kan}\ \bibnamefont
  {Zhao}}, \bibinfo {author} {\bibfnamefont {Hao}\ \bibnamefont {Deng}},
  \bibinfo {author} {\bibfnamefont {Hua}\ \bibnamefont {Chen}}, \bibinfo
  {author} {\bibfnamefont {Kate~A.}\ \bibnamefont {Ross}}, \bibinfo {author}
  {\bibfnamefont {Vaclav}\ \bibnamefont {Petříček}}, \bibinfo {author}
  {\bibfnamefont {Gerrit}\ \bibnamefont {Günther}}, \bibinfo {author}
  {\bibfnamefont {Margarita}\ \bibnamefont {Russina}}, \bibinfo {author}
  {\bibfnamefont {Vladimir}\ \bibnamefont {Hutanu}}, \ and\ \bibinfo {author}
  {\bibfnamefont {Philipp}\ \bibnamefont {Gegenwart}},\ }\bibfield  {title}
  {\enquote {\bibinfo {title} {Realization of the kagome spin ice state in a
  frustrated intermetallic compound},}\ }\href {\doibase
  10.1126/science.aaw1666} {\bibfield  {journal} {\bibinfo  {journal}
  {Science}\ }\textbf {\bibinfo {volume} {367}},\ \bibinfo {pages} {1218--1223}
  (\bibinfo {year} {2020})}\BibitemShut {NoStop}%
\bibitem [{\citenamefont {Gibson}\ \emph {et~al.}(1996)\citenamefont {Gibson},
  \citenamefont {Pöttgen}, \citenamefont {Kremer}, \citenamefont {Simon},\
  and\ \citenamefont {Ziebeck}}]{GIBSON199634}%
  \BibitemOpen
  \bibfield  {author} {\bibinfo {author} {\bibfnamefont {Brendan}\ \bibnamefont
  {Gibson}}, \bibinfo {author} {\bibfnamefont {Rainer}\ \bibnamefont
  {Pöttgen}}, \bibinfo {author} {\bibfnamefont {Reinhard~K.}\ \bibnamefont
  {Kremer}}, \bibinfo {author} {\bibfnamefont {Arndt}\ \bibnamefont {Simon}}, \
  and\ \bibinfo {author} {\bibfnamefont {Kurt~R.A.}\ \bibnamefont {Ziebeck}},\
  }\bibfield  {title} {\enquote {\bibinfo {title} {{Ternary germanides LnAgGe
  (Ln = Y, Sm, Gd-Lu) with ordered Fe$_2$P-type structure}},}\ }\href {\doibase
  https://doi.org/10.1016/0925-8388(96)02201-3} {\bibfield  {journal} {\bibinfo
   {journal} {Journal of Alloys and Compounds}\ }\textbf {\bibinfo {volume}
  {239}},\ \bibinfo {pages} {34--40} (\bibinfo {year} {1996})}\BibitemShut
  {NoStop}%
\bibitem [{\citenamefont {Baran}\ \emph {et~al.}(1998)\citenamefont {Baran},
  \citenamefont {Hofmann}, \citenamefont {Leciejewicz}, \citenamefont {Penc},
  \citenamefont {Ślaski},\ and\ \citenamefont {Szytuła}}]{BARAN_1998}%
  \BibitemOpen
  \bibfield  {author} {\bibinfo {author} {\bibfnamefont {S.}~\bibnamefont
  {Baran}}, \bibinfo {author} {\bibfnamefont {M.}~\bibnamefont {Hofmann}},
  \bibinfo {author} {\bibfnamefont {J.}~\bibnamefont {Leciejewicz}}, \bibinfo
  {author} {\bibfnamefont {B.}~\bibnamefont {Penc}}, \bibinfo {author}
  {\bibfnamefont {M.}~\bibnamefont {Ślaski}}, \ and\ \bibinfo {author}
  {\bibfnamefont {A.}~\bibnamefont {Szytuła}},\ }\bibfield  {title} {\enquote
  {\bibinfo {title} {{Magnetic order in RAgGe (R=Gd-Er) intermetallic
  compounds}},}\ }\href {\doibase
  https://doi.org/10.1016/S0925-8388(98)00721-X} {\bibfield  {journal}
  {\bibinfo  {journal} {Journal of Alloys and Compounds}\ }\textbf {\bibinfo
  {volume} {281}},\ \bibinfo {pages} {92--98} (\bibinfo {year}
  {1998})}\BibitemShut {NoStop}%
\bibitem [{\citenamefont {Li}\ \emph {et~al.}(2022)\citenamefont {Li},
  \citenamefont {Huang}, \citenamefont {Yue}, \citenamefont {Guang},
  \citenamefont {Xia}, \citenamefont {Wang}, \citenamefont {Li}, \citenamefont
  {Zhao}, \citenamefont {Zhou},\ and\ \citenamefont {Sun}}]{Li_2022_PhysRevB}%
  \BibitemOpen
  \bibfield  {author} {\bibinfo {author} {\bibfnamefont {N.}~\bibnamefont
  {Li}}, \bibinfo {author} {\bibfnamefont {Q.}~\bibnamefont {Huang}}, \bibinfo
  {author} {\bibfnamefont {X.~Y.}\ \bibnamefont {Yue}}, \bibinfo {author}
  {\bibfnamefont {S.~K.}\ \bibnamefont {Guang}}, \bibinfo {author}
  {\bibfnamefont {K.}~\bibnamefont {Xia}}, \bibinfo {author} {\bibfnamefont
  {Y.~Y.}\ \bibnamefont {Wang}}, \bibinfo {author} {\bibfnamefont {Q.~J.}\
  \bibnamefont {Li}}, \bibinfo {author} {\bibfnamefont {X.}~\bibnamefont
  {Zhao}}, \bibinfo {author} {\bibfnamefont {H.~D.}\ \bibnamefont {Zhou}}, \
  and\ \bibinfo {author} {\bibfnamefont {X.~F.}\ \bibnamefont {Sun}},\
  }\bibfield  {title} {\enquote {\bibinfo {title} {Low-temperature transport
  properties of the intermetallic compound hoagge with a kagome spin-ice
  state},}\ }\href {\doibase 10.1103/PhysRevB.106.014416} {\bibfield  {journal}
  {\bibinfo  {journal} {Phys. Rev. B}\ }\textbf {\bibinfo {volume} {106}},\
  \bibinfo {pages} {014416} (\bibinfo {year} {2022})}\BibitemShut {NoStop}%
\bibitem [{\citenamefont {Zhao}\ \emph {et~al.}(2024)\citenamefont {Zhao},
  \citenamefont {Tokiwa}, \citenamefont {Chen},\ and\ \citenamefont
  {Gegenwart}}]{Zhao_2024_Nature_Phy}%
  \BibitemOpen
  \bibfield  {author} {\bibinfo {author} {\bibfnamefont {K.}~\bibnamefont
  {Zhao}}, \bibinfo {author} {\bibfnamefont {Y.}~\bibnamefont {Tokiwa}},
  \bibinfo {author} {\bibfnamefont {H.}~\bibnamefont {Chen}}, \ and\ \bibinfo
  {author} {\bibfnamefont {P.}~\bibnamefont {Gegenwart}},\ }\bibfield  {title}
  {\enquote {\bibinfo {title} {Discrete degeneracies distinguished by the
  anomalous hall effect in a metallic kagome ice compound},}\ }\href {\doibase
  10.1038/s41567-023-02307-w} {\bibfield  {journal} {\bibinfo  {journal}
  {Nature Physics}\ }\textbf {\bibinfo {volume} {20}},\ \bibinfo {pages}
  {442--449} (\bibinfo {year} {2024})}\BibitemShut {NoStop}%
\bibitem [{\citenamefont {Deng}\ \emph {et~al.}(2024)\citenamefont {Deng},
  \citenamefont {Yang}, \citenamefont {Liu}, \citenamefont {Liu}, \citenamefont
  {Zhao}, \citenamefont {Wang}, \citenamefont {Li}, \citenamefont {Song},
  \citenamefont {Neupert}, \citenamefont {Liu}, \citenamefont {Shao},
  \citenamefont {Zhao}, \citenamefont {Xu}, \citenamefont {Deng}, \citenamefont
  {Huang}, \citenamefont {Zhao}, \citenamefont {Zhang}, \citenamefont {Mei},
  \citenamefont {Wu}, \citenamefont {He}, \citenamefont {Liu}, \citenamefont
  {Liu},\ and\ \citenamefont {Yin}}]{Deng_2024PhysRevLett}%
  \BibitemOpen
  \bibfield  {author} {\bibinfo {author} {\bibfnamefont {Hanbin}\ \bibnamefont
  {Deng}}, \bibinfo {author} {\bibfnamefont {Tianyu}\ \bibnamefont {Yang}},
  \bibinfo {author} {\bibfnamefont {Guowei}\ \bibnamefont {Liu}}, \bibinfo
  {author} {\bibfnamefont {Lu}~\bibnamefont {Liu}}, \bibinfo {author}
  {\bibfnamefont {Lingxiao}\ \bibnamefont {Zhao}}, \bibinfo {author}
  {\bibfnamefont {Wu}~\bibnamefont {Wang}}, \bibinfo {author} {\bibfnamefont
  {Tiantian}\ \bibnamefont {Li}}, \bibinfo {author} {\bibfnamefont {Wei}\
  \bibnamefont {Song}}, \bibinfo {author} {\bibfnamefont {Titus}\ \bibnamefont
  {Neupert}}, \bibinfo {author} {\bibfnamefont {Xiang-Rui}\ \bibnamefont
  {Liu}}, \bibinfo {author} {\bibfnamefont {Jifeng}\ \bibnamefont {Shao}},
  \bibinfo {author} {\bibfnamefont {Y.~Y.}\ \bibnamefont {Zhao}}, \bibinfo
  {author} {\bibfnamefont {Nan}\ \bibnamefont {Xu}}, \bibinfo {author}
  {\bibfnamefont {Hao}\ \bibnamefont {Deng}}, \bibinfo {author} {\bibfnamefont
  {Li}~\bibnamefont {Huang}}, \bibinfo {author} {\bibfnamefont {Yue}\
  \bibnamefont {Zhao}}, \bibinfo {author} {\bibfnamefont {Liyuan}\ \bibnamefont
  {Zhang}}, \bibinfo {author} {\bibfnamefont {Jia-Wei}\ \bibnamefont {Mei}},
  \bibinfo {author} {\bibfnamefont {Liusuo}\ \bibnamefont {Wu}}, \bibinfo
  {author} {\bibfnamefont {Jiaqing}\ \bibnamefont {He}}, \bibinfo {author}
  {\bibfnamefont {Qihang}\ \bibnamefont {Liu}}, \bibinfo {author}
  {\bibfnamefont {Chang}\ \bibnamefont {Liu}}, \ and\ \bibinfo {author}
  {\bibfnamefont {Jia-Xin}\ \bibnamefont {Yin}},\ }\bibfield  {title} {\enquote
  {\bibinfo {title} {Local excitation of kagome spin ice magnetism seen by
  scanning tunneling microscopy},}\ }\href {\doibase
  10.1103/PhysRevLett.133.046503} {\bibfield  {journal} {\bibinfo  {journal}
  {Phys. Rev. Lett.}\ }\textbf {\bibinfo {volume} {133}},\ \bibinfo {pages}
  {046503} (\bibinfo {year} {2024})}\BibitemShut {NoStop}%
\bibitem [{\citenamefont {Giannozzi}\ \emph {et~al.}(2009)\citenamefont
  {Giannozzi}, \citenamefont {Baroni}, \citenamefont {Bonini}, \citenamefont
  {Calandra}, \citenamefont {Car}, \citenamefont {Cavazzoni}, \citenamefont
  {Ceresoli}, \citenamefont {Chiarotti}, \citenamefont {Cococcioni},
  \citenamefont {Dabo}, \citenamefont {Corso}, \citenamefont {de~Gironcoli},
  \citenamefont {Fabris}, \citenamefont {Fratesi}, \citenamefont {Gebauer},
  \citenamefont {Gerstmann}, \citenamefont {Gougoussis}, \citenamefont
  {Kokalj}, \citenamefont {Lazzeri}, \citenamefont {Martin-Samos},
  \citenamefont {Marzari}, \citenamefont {Mauri}, \citenamefont {Mazzarello},
  \citenamefont {Paolini}, \citenamefont {Pasquarello}, \citenamefont
  {Paulatto}, \citenamefont {Sbraccia}, \citenamefont {Scandolo}, \citenamefont
  {Sclauzero}, \citenamefont {Seitsonen}, \citenamefont {Smogunov},
  \citenamefont {Umari},\ and\ \citenamefont
  {Wentzcovitch}}]{Giannozzi_2009_JPCM}%
  \BibitemOpen
  \bibfield  {author} {\bibinfo {author} {\bibfnamefont {Paolo}\ \bibnamefont
  {Giannozzi}}, \bibinfo {author} {\bibfnamefont {Stefano}\ \bibnamefont
  {Baroni}}, \bibinfo {author} {\bibfnamefont {Nicola}\ \bibnamefont {Bonini}},
  \bibinfo {author} {\bibfnamefont {Matteo}\ \bibnamefont {Calandra}}, \bibinfo
  {author} {\bibfnamefont {Roberto}\ \bibnamefont {Car}}, \bibinfo {author}
  {\bibfnamefont {Carlo}\ \bibnamefont {Cavazzoni}}, \bibinfo {author}
  {\bibfnamefont {Davide}\ \bibnamefont {Ceresoli}}, \bibinfo {author}
  {\bibfnamefont {Guido~L}\ \bibnamefont {Chiarotti}}, \bibinfo {author}
  {\bibfnamefont {Matteo}\ \bibnamefont {Cococcioni}}, \bibinfo {author}
  {\bibfnamefont {Ismaila}\ \bibnamefont {Dabo}}, \bibinfo {author}
  {\bibfnamefont {Andrea~Dal}\ \bibnamefont {Corso}}, \bibinfo {author}
  {\bibfnamefont {Stefano}\ \bibnamefont {de~Gironcoli}}, \bibinfo {author}
  {\bibfnamefont {Stefano}\ \bibnamefont {Fabris}}, \bibinfo {author}
  {\bibfnamefont {Guido}\ \bibnamefont {Fratesi}}, \bibinfo {author}
  {\bibfnamefont {Ralph}\ \bibnamefont {Gebauer}}, \bibinfo {author}
  {\bibfnamefont {Uwe}\ \bibnamefont {Gerstmann}}, \bibinfo {author}
  {\bibfnamefont {Christos}\ \bibnamefont {Gougoussis}}, \bibinfo {author}
  {\bibfnamefont {Anton}\ \bibnamefont {Kokalj}}, \bibinfo {author}
  {\bibfnamefont {Michele}\ \bibnamefont {Lazzeri}}, \bibinfo {author}
  {\bibfnamefont {Layla}\ \bibnamefont {Martin-Samos}}, \bibinfo {author}
  {\bibfnamefont {Nicola}\ \bibnamefont {Marzari}}, \bibinfo {author}
  {\bibfnamefont {Francesco}\ \bibnamefont {Mauri}}, \bibinfo {author}
  {\bibfnamefont {Riccardo}\ \bibnamefont {Mazzarello}}, \bibinfo {author}
  {\bibfnamefont {Stefano}\ \bibnamefont {Paolini}}, \bibinfo {author}
  {\bibfnamefont {Alfredo}\ \bibnamefont {Pasquarello}}, \bibinfo {author}
  {\bibfnamefont {Lorenzo}\ \bibnamefont {Paulatto}}, \bibinfo {author}
  {\bibfnamefont {Carlo}\ \bibnamefont {Sbraccia}}, \bibinfo {author}
  {\bibfnamefont {Sandro}\ \bibnamefont {Scandolo}}, \bibinfo {author}
  {\bibfnamefont {Gabriele}\ \bibnamefont {Sclauzero}}, \bibinfo {author}
  {\bibfnamefont {Ari~P}\ \bibnamefont {Seitsonen}}, \bibinfo {author}
  {\bibfnamefont {Alexander}\ \bibnamefont {Smogunov}}, \bibinfo {author}
  {\bibfnamefont {Paolo}\ \bibnamefont {Umari}}, \ and\ \bibinfo {author}
  {\bibfnamefont {Renata~M}\ \bibnamefont {Wentzcovitch}},\ }\bibfield  {title}
  {\enquote {\bibinfo {title} {{QUANTUM ESPRESSO: a modular and open-source
  software project for quantum simulations of materials}},}\ }\href {\doibase
  10.1088/0953-8984/21/39/395502} {\bibfield  {journal} {\bibinfo  {journal}
  {Journal of Physics: Condensed Matter}\ }\textbf {\bibinfo {volume} {21}},\
  \bibinfo {pages} {395502} (\bibinfo {year} {2009})}\BibitemShut {NoStop}%
\bibitem [{\citenamefont {Giannozzi}\ \emph {et~al.}(2017)\citenamefont
  {Giannozzi}, \citenamefont {Andreussi}, \citenamefont {Brumme}, \citenamefont
  {Bunau}, \citenamefont {Nardelli}, \citenamefont {Calandra}, \citenamefont
  {Car}, \citenamefont {Cavazzoni}, \citenamefont {Ceresoli}, \citenamefont
  {Cococcioni}, \citenamefont {Colonna}, \citenamefont {Carnimeo},
  \citenamefont {Corso}, \citenamefont {de~Gironcoli}, \citenamefont {Delugas},
  \citenamefont {DiStasio}, \citenamefont {Ferretti}, \citenamefont {Floris},
  \citenamefont {Fratesi}, \citenamefont {Fugallo}, \citenamefont {Gebauer},
  \citenamefont {Gerstmann}, \citenamefont {Giustino}, \citenamefont {Gorni},
  \citenamefont {Jia}, \citenamefont {Kawamura}, \citenamefont {Ko},
  \citenamefont {Kokalj}, \citenamefont {Küçükbenli}, \citenamefont
  {Lazzeri}, \citenamefont {Marsili}, \citenamefont {Marzari}, \citenamefont
  {Mauri}, \citenamefont {Nguyen}, \citenamefont {Nguyen}, \citenamefont {de-la
  Roza}, \citenamefont {Paulatto}, \citenamefont {Poncé}, \citenamefont
  {Rocca}, \citenamefont {Sabatini}, \citenamefont {Santra}, \citenamefont
  {Schlipf}, \citenamefont {Seitsonen}, \citenamefont {Smogunov}, \citenamefont
  {Timrov}, \citenamefont {Thonhauser}, \citenamefont {Umari}, \citenamefont
  {Vast}, \citenamefont {Wu},\ and\ \citenamefont {Baroni}}]{Giannozzi_2017}%
  \BibitemOpen
  \bibfield  {author} {\bibinfo {author} {\bibfnamefont {P}~\bibnamefont
  {Giannozzi}}, \bibinfo {author} {\bibfnamefont {O}~\bibnamefont {Andreussi}},
  \bibinfo {author} {\bibfnamefont {T}~\bibnamefont {Brumme}}, \bibinfo
  {author} {\bibfnamefont {O}~\bibnamefont {Bunau}}, \bibinfo {author}
  {\bibfnamefont {M~Buongiorno}\ \bibnamefont {Nardelli}}, \bibinfo {author}
  {\bibfnamefont {M}~\bibnamefont {Calandra}}, \bibinfo {author} {\bibfnamefont
  {R}~\bibnamefont {Car}}, \bibinfo {author} {\bibfnamefont {C}~\bibnamefont
  {Cavazzoni}}, \bibinfo {author} {\bibfnamefont {D}~\bibnamefont {Ceresoli}},
  \bibinfo {author} {\bibfnamefont {M}~\bibnamefont {Cococcioni}}, \bibinfo
  {author} {\bibfnamefont {N}~\bibnamefont {Colonna}}, \bibinfo {author}
  {\bibfnamefont {I}~\bibnamefont {Carnimeo}}, \bibinfo {author} {\bibfnamefont
  {A~Dal}\ \bibnamefont {Corso}}, \bibinfo {author} {\bibfnamefont
  {S}~\bibnamefont {de~Gironcoli}}, \bibinfo {author} {\bibfnamefont
  {P}~\bibnamefont {Delugas}}, \bibinfo {author} {\bibfnamefont {R~A}\
  \bibnamefont {DiStasio}}, \bibinfo {author} {\bibfnamefont {A}~\bibnamefont
  {Ferretti}}, \bibinfo {author} {\bibfnamefont {A}~\bibnamefont {Floris}},
  \bibinfo {author} {\bibfnamefont {G}~\bibnamefont {Fratesi}}, \bibinfo
  {author} {\bibfnamefont {G}~\bibnamefont {Fugallo}}, \bibinfo {author}
  {\bibfnamefont {R}~\bibnamefont {Gebauer}}, \bibinfo {author} {\bibfnamefont
  {U}~\bibnamefont {Gerstmann}}, \bibinfo {author} {\bibfnamefont
  {F}~\bibnamefont {Giustino}}, \bibinfo {author} {\bibfnamefont
  {T}~\bibnamefont {Gorni}}, \bibinfo {author} {\bibfnamefont {J}~\bibnamefont
  {Jia}}, \bibinfo {author} {\bibfnamefont {M}~\bibnamefont {Kawamura}},
  \bibinfo {author} {\bibfnamefont {H-Y}\ \bibnamefont {Ko}}, \bibinfo {author}
  {\bibfnamefont {A}~\bibnamefont {Kokalj}}, \bibinfo {author} {\bibfnamefont
  {E}~\bibnamefont {Küçükbenli}}, \bibinfo {author} {\bibfnamefont
  {M}~\bibnamefont {Lazzeri}}, \bibinfo {author} {\bibfnamefont
  {M}~\bibnamefont {Marsili}}, \bibinfo {author} {\bibfnamefont
  {N}~\bibnamefont {Marzari}}, \bibinfo {author} {\bibfnamefont
  {F}~\bibnamefont {Mauri}}, \bibinfo {author} {\bibfnamefont {N~L}\
  \bibnamefont {Nguyen}}, \bibinfo {author} {\bibfnamefont {H-V}\ \bibnamefont
  {Nguyen}}, \bibinfo {author} {\bibfnamefont {A~Otero}\ \bibnamefont {de-la
  Roza}}, \bibinfo {author} {\bibfnamefont {L}~\bibnamefont {Paulatto}},
  \bibinfo {author} {\bibfnamefont {S}~\bibnamefont {Poncé}}, \bibinfo
  {author} {\bibfnamefont {D}~\bibnamefont {Rocca}}, \bibinfo {author}
  {\bibfnamefont {R}~\bibnamefont {Sabatini}}, \bibinfo {author} {\bibfnamefont
  {B}~\bibnamefont {Santra}}, \bibinfo {author} {\bibfnamefont {M}~\bibnamefont
  {Schlipf}}, \bibinfo {author} {\bibfnamefont {A~P}\ \bibnamefont
  {Seitsonen}}, \bibinfo {author} {\bibfnamefont {A}~\bibnamefont {Smogunov}},
  \bibinfo {author} {\bibfnamefont {I}~\bibnamefont {Timrov}}, \bibinfo
  {author} {\bibfnamefont {T}~\bibnamefont {Thonhauser}}, \bibinfo {author}
  {\bibfnamefont {P}~\bibnamefont {Umari}}, \bibinfo {author} {\bibfnamefont
  {N}~\bibnamefont {Vast}}, \bibinfo {author} {\bibfnamefont {X}~\bibnamefont
  {Wu}}, \ and\ \bibinfo {author} {\bibfnamefont {S}~\bibnamefont {Baroni}},\
  }\bibfield  {title} {\enquote {\bibinfo {title} {{Advanced capabilities for
  materials modelling with Quantum ESPRESSO}},}\ }\href {\doibase
  10.1088/1361-648X/aa8f79} {\bibfield  {journal} {\bibinfo  {journal} {Journal
  of Physics: Condensed Matter}\ }\textbf {\bibinfo {volume} {29}},\ \bibinfo
  {pages} {465901} (\bibinfo {year} {2017})}\BibitemShut {NoStop}%
\bibitem [{\citenamefont {Perdew}\ \emph {et~al.}(1996)\citenamefont {Perdew},
  \citenamefont {Burke},\ and\ \citenamefont
  {Ernzerhof}}]{Perdew1996PhysRevLett}%
  \BibitemOpen
  \bibfield  {author} {\bibinfo {author} {\bibfnamefont {John~P.}\ \bibnamefont
  {Perdew}}, \bibinfo {author} {\bibfnamefont {Kieron}\ \bibnamefont {Burke}},
  \ and\ \bibinfo {author} {\bibfnamefont {Matthias}\ \bibnamefont
  {Ernzerhof}},\ }\bibfield  {title} {\enquote {\bibinfo {title} {Generalized
  gradient approximation made simple},}\ }\href {\doibase
  10.1103/PhysRevLett.77.3865} {\bibfield  {journal} {\bibinfo  {journal}
  {Phys. Rev. Lett.}\ }\textbf {\bibinfo {volume} {77}},\ \bibinfo {pages}
  {3865--3868} (\bibinfo {year} {1996})}\BibitemShut {NoStop}%
\bibitem [{\citenamefont {{van Setten}}\ \emph {et~al.}(2018)\citenamefont
  {{van Setten}}, \citenamefont {Giantomassi}, \citenamefont {Bousquet},
  \citenamefont {Verstraete}, \citenamefont {Hamann}, \citenamefont {Gonze},\
  and\ \citenamefont {Rignanese}}]{VANSETTEN201839}%
  \BibitemOpen
  \bibfield  {author} {\bibinfo {author} {\bibfnamefont {M.J.}\ \bibnamefont
  {{van Setten}}}, \bibinfo {author} {\bibfnamefont {M.}~\bibnamefont
  {Giantomassi}}, \bibinfo {author} {\bibfnamefont {E.}~\bibnamefont
  {Bousquet}}, \bibinfo {author} {\bibfnamefont {M.J.}\ \bibnamefont
  {Verstraete}}, \bibinfo {author} {\bibfnamefont {D.R.}\ \bibnamefont
  {Hamann}}, \bibinfo {author} {\bibfnamefont {X.}~\bibnamefont {Gonze}}, \
  and\ \bibinfo {author} {\bibfnamefont {G.-M.}\ \bibnamefont {Rignanese}},\
  }\bibfield  {title} {\enquote {\bibinfo {title} {{The PseudoDojo: Training
  and grading a 85 element optimized norm-conserving pseudopotential table}},}\
  }\href {\doibase https://doi.org/10.1016/j.cpc.2018.01.012} {\bibfield
  {journal} {\bibinfo  {journal} {Computer Physics Communications}\ }\textbf
  {\bibinfo {volume} {226}},\ \bibinfo {pages} {39--54} (\bibinfo {year}
  {2018})}\BibitemShut {NoStop}%
\bibitem [{\citenamefont {Togo}\ and\ \citenamefont
  {Tanaka}(2015)}]{TOGO20151}%
  \BibitemOpen
  \bibfield  {author} {\bibinfo {author} {\bibfnamefont {Atsushi}\ \bibnamefont
  {Togo}}\ and\ \bibinfo {author} {\bibfnamefont {Isao}\ \bibnamefont
  {Tanaka}},\ }\bibfield  {title} {\enquote {\bibinfo {title} {First principles
  phonon calculations in materials science},}\ }\href {\doibase
  https://doi.org/10.1016/j.scriptamat.2015.07.021} {\bibfield  {journal}
  {\bibinfo  {journal} {Scripta Materialia}\ }\textbf {\bibinfo {volume}
  {108}},\ \bibinfo {pages} {1--5} (\bibinfo {year} {2015})}\BibitemShut
  {NoStop}%
\bibitem [{\citenamefont {Kroumova}\ \emph {et~al.}(2003)\citenamefont
  {Kroumova}, \citenamefont {Aroyo}, \citenamefont {Perez-Mato}, \citenamefont
  {Kirov}, \citenamefont {Capillas}, \citenamefont {Ivantchev},\ and\
  \citenamefont {Wondratschek}}]{Bilbao_1}%
  \BibitemOpen
  \bibfield  {author} {\bibinfo {author} {\bibfnamefont {E.}~\bibnamefont
  {Kroumova}}, \bibinfo {author} {\bibfnamefont {M.~I.}\ \bibnamefont {Aroyo}},
  \bibinfo {author} {\bibfnamefont {J.~M.}\ \bibnamefont {Perez-Mato}},
  \bibinfo {author} {\bibfnamefont {A.}~\bibnamefont {Kirov}}, \bibinfo
  {author} {\bibfnamefont {C.}~\bibnamefont {Capillas}}, \bibinfo {author}
  {\bibfnamefont {S.}~\bibnamefont {Ivantchev}}, \ and\ \bibinfo {author}
  {\bibfnamefont {H.}~\bibnamefont {Wondratschek}},\ }\bibfield  {title}
  {\enquote {\bibinfo {title} {Bilbao crystallographic server: Useful databases
  and tools for phase-transition studies},}\ }\href
  {http://dx.doi.org/10.1080/0141159031000076110} {\bibfield  {journal}
  {\bibinfo  {journal} {Phase Transit.}\ }\textbf {\bibinfo {volume} {76}},\
  \bibinfo {pages} {155--170} (\bibinfo {year} {2003})}\BibitemShut {NoStop}%
\bibitem [{\citenamefont {Aroyo}\ \emph {et~al.}(2011)\citenamefont {Aroyo},
  \citenamefont {Perez-Mato}, \citenamefont {Orobengoa}, \citenamefont {Tasci},
  \citenamefont {De~La~Flor},\ and\ \citenamefont {Kirov}}]{Bilbao_4}%
  \BibitemOpen
  \bibfield  {author} {\bibinfo {author} {\bibfnamefont {M.~I.}\ \bibnamefont
  {Aroyo}}, \bibinfo {author} {\bibfnamefont {J.~M.}\ \bibnamefont
  {Perez-Mato}}, \bibinfo {author} {\bibfnamefont {D.}~\bibnamefont
  {Orobengoa}}, \bibinfo {author} {\bibfnamefont {E.}~\bibnamefont {Tasci}},
  \bibinfo {author} {\bibfnamefont {G.}~\bibnamefont {De~La~Flor}}, \ and\
  \bibinfo {author} {\bibfnamefont {A.}~\bibnamefont {Kirov}},\ }\bibfield
  {title} {\enquote {\bibinfo {title} {Crystallography online: Bilbao
  crystallographic server},}\ }\href
  {https://www.scopus.com/record/display.uri?eid=2-s2.0-80955140447&origin=AuthorNamesList&txGid=dGz08KXSuEoo96iJokT3Rjd\%3a3}
  {\bibfield  {journal} {\bibinfo  {journal} {Bulg. Chem. Commun}\ }\textbf
  {\bibinfo {volume} {43}},\ \bibinfo {pages} {183--97} (\bibinfo {year}
  {2011})}\BibitemShut {NoStop}%
\bibitem [{\citenamefont {Hatch}\ and\ \citenamefont
  {Stokes}(2003)}]{Hatch2003}%
  \BibitemOpen
  \bibfield  {author} {\bibinfo {author} {\bibfnamefont {Dorian~M}\
  \bibnamefont {Hatch}}\ and\ \bibinfo {author} {\bibfnamefont {Harold~T}\
  \bibnamefont {Stokes}},\ }\bibfield  {title} {\enquote {\bibinfo {title}
  {Invariants: program for obtaining a list of invariant polynomials of the
  order-parameter components associated with irreducible representations of a
  space group},}\ }\href@noop {} {\bibfield  {journal} {\bibinfo  {journal}
  {Journal of applied crystallography}\ }\textbf {\bibinfo {volume} {36}},\
  \bibinfo {pages} {951--952} (\bibinfo {year} {2003})}\BibitemShut {NoStop}%
\bibitem [{\citenamefont {Cracknell}\ \emph {et~al.}(1979)\citenamefont
  {Cracknell}, \citenamefont {Davies}, \citenamefont {Miller},\ and\
  \citenamefont {Love}}]{cracknell1979general}%
  \BibitemOpen
  \bibfield  {author} {\bibinfo {author} {\bibfnamefont {A.~P.}\ \bibnamefont
  {Cracknell}}, \bibinfo {author} {\bibfnamefont {B.~L.}\ \bibnamefont
  {Davies}}, \bibinfo {author} {\bibfnamefont {S.~C.}\ \bibnamefont {Miller}},
  \ and\ \bibinfo {author} {\bibfnamefont {W.~F.}\ \bibnamefont {Love}},\
  }\bibfield  {title} {\enquote {\bibinfo {title} {{Kronecker Product Tables.
  Vol. 1. General introduction and Tables of irreducible representations of
  space groups}},}\ \ }(\bibinfo  {publisher} {IFI/Plenum, New York},\ \bibinfo
  {year} {1979})\BibitemShut {NoStop}%
\bibitem [{\citenamefont {Fano}(1961)}]{Fano_1961PhysRev}%
  \BibitemOpen
  \bibfield  {author} {\bibinfo {author} {\bibfnamefont {U.}~\bibnamefont
  {Fano}},\ }\bibfield  {title} {\enquote {\bibinfo {title} {Effects of
  configuration interaction on intensities and phase shifts},}\ }\href
  {\doibase 10.1103/PhysRev.124.1866} {\bibfield  {journal} {\bibinfo
  {journal} {Phys. Rev.}\ }\textbf {\bibinfo {volume} {124}},\ \bibinfo {pages}
  {1866--1878} (\bibinfo {year} {1961})}\BibitemShut {NoStop}%
\bibitem [{\citenamefont {Wu}\ \emph {et~al.}(2020)\citenamefont {Wu},
  \citenamefont {Zhang}, \citenamefont {Li}, \citenamefont {Cao}, \citenamefont
  {Kung}, \citenamefont {Sefat}, \citenamefont {Ding}, \citenamefont
  {Richard},\ and\ \citenamefont {Blumberg}}]{Wu_2020PhysRevB}%
  \BibitemOpen
  \bibfield  {author} {\bibinfo {author} {\bibfnamefont {S.-F.}\ \bibnamefont
  {Wu}}, \bibinfo {author} {\bibfnamefont {W.-L.}\ \bibnamefont {Zhang}},
  \bibinfo {author} {\bibfnamefont {L.}~\bibnamefont {Li}}, \bibinfo {author}
  {\bibfnamefont {H.-B.}\ \bibnamefont {Cao}}, \bibinfo {author} {\bibfnamefont
  {H.-H.}\ \bibnamefont {Kung}}, \bibinfo {author} {\bibfnamefont {A.~S.}\
  \bibnamefont {Sefat}}, \bibinfo {author} {\bibfnamefont {H.}~\bibnamefont
  {Ding}}, \bibinfo {author} {\bibfnamefont {P.}~\bibnamefont {Richard}}, \
  and\ \bibinfo {author} {\bibfnamefont {G.}~\bibnamefont {Blumberg}},\
  }\bibfield  {title} {\enquote {\bibinfo {title} {{Coupling of fully symmetric
  As phonon to magnetism in
  $\mathrm{Ba}{({\mathrm{Fe}}_{1\ensuremath{-}x}{\mathrm{Au}}_{x})}_{2}{\mathrm{As}}_{2}$}},}\
  }\href {\doibase 10.1103/PhysRevB.102.014501} {\bibfield  {journal} {\bibinfo
   {journal} {Phys. Rev. B}\ }\textbf {\bibinfo {volume} {102}},\ \bibinfo
  {pages} {014501} (\bibinfo {year} {2020})}\BibitemShut {NoStop}%
\bibitem [{\citenamefont {Lockwood}\ and\ \citenamefont
  {Cottam}(1988)}]{Lockwood_1988JAP}%
  \BibitemOpen
  \bibfield  {author} {\bibinfo {author} {\bibfnamefont {D.~J.}\ \bibnamefont
  {Lockwood}}\ and\ \bibinfo {author} {\bibfnamefont {M.~G.}\ \bibnamefont
  {Cottam}},\ }\bibfield  {title} {\enquote {\bibinfo {title} {{The spin-phonon
  interaction in FeF$_2$ and MnF$_2$ studied by Raman spectroscopy}},}\ }\href
  {\doibase 10.1063/1.342186} {\bibfield  {journal} {\bibinfo  {journal}
  {Journal of Applied Physics}\ }\textbf {\bibinfo {volume} {64}},\ \bibinfo
  {pages} {5876--5878} (\bibinfo {year} {1988})}\BibitemShut {NoStop}%
\bibitem [{\citenamefont {Lockwood}(2002)}]{Lockwood_2002LTP}%
  \BibitemOpen
  \bibfield  {author} {\bibinfo {author} {\bibfnamefont {D.~J.}\ \bibnamefont
  {Lockwood}},\ }\bibfield  {title} {\enquote {\bibinfo {title} {{Spin–phonon
  interaction and mode softening in NiF$_2$}},}\ }\href {\doibase
  10.1063/1.1496657} {\bibfield  {journal} {\bibinfo  {journal} {Low
  Temperature Physics}\ }\textbf {\bibinfo {volume} {28}},\ \bibinfo {pages}
  {505--509} (\bibinfo {year} {2002})}\BibitemShut {NoStop}%
\bibitem [{\citenamefont {Zhang}\ \emph {et~al.}(2008)\citenamefont {Zhang},
  \citenamefont {An}, \citenamefont {Yuan}, \citenamefont {Wu}, \citenamefont
  {Wu}, \citenamefont {Luo}, \citenamefont {Wang}, \citenamefont {Bao},\ and\
  \citenamefont {Wang}}]{Zhang_2008PhysRevB}%
  \BibitemOpen
  \bibfield  {author} {\bibinfo {author} {\bibfnamefont {Qingming}\
  \bibnamefont {Zhang}}, \bibinfo {author} {\bibfnamefont {Ming}\ \bibnamefont
  {An}}, \bibinfo {author} {\bibfnamefont {Shikui}\ \bibnamefont {Yuan}},
  \bibinfo {author} {\bibfnamefont {Yong}\ \bibnamefont {Wu}}, \bibinfo
  {author} {\bibfnamefont {Dong}\ \bibnamefont {Wu}}, \bibinfo {author}
  {\bibfnamefont {Jianlin}\ \bibnamefont {Luo}}, \bibinfo {author}
  {\bibfnamefont {Nanlin}\ \bibnamefont {Wang}}, \bibinfo {author}
  {\bibfnamefont {Wei}\ \bibnamefont {Bao}}, \ and\ \bibinfo {author}
  {\bibfnamefont {Yening}\ \bibnamefont {Wang}},\ }\bibfield  {title} {\enquote
  {\bibinfo {title} {{Phonon softening and forbidden mode in
  ${\mathrm{Na}}_{0.5}\mathrm{Co}{\mathrm{O}}_{2}$ observed by Raman
  scattering}},}\ }\href {\doibase 10.1103/PhysRevB.77.045110} {\bibfield
  {journal} {\bibinfo  {journal} {Phys. Rev. B}\ }\textbf {\bibinfo {volume}
  {77}},\ \bibinfo {pages} {045110} (\bibinfo {year} {2008})}\BibitemShut
  {NoStop}%
\bibitem [{\citenamefont {Granado}\ \emph {et~al.}(1999)\citenamefont
  {Granado}, \citenamefont {Garc\'{\i}a}, \citenamefont {Sanjurjo},
  \citenamefont {Rettori}, \citenamefont {Torriani}, \citenamefont {Prado},
  \citenamefont {S\'anchez}, \citenamefont {Caneiro},\ and\ \citenamefont
  {Oseroff}}]{Granado_1999PhysRevB}%
  \BibitemOpen
  \bibfield  {author} {\bibinfo {author} {\bibfnamefont {E.}~\bibnamefont
  {Granado}}, \bibinfo {author} {\bibfnamefont {A.}~\bibnamefont
  {Garc\'{\i}a}}, \bibinfo {author} {\bibfnamefont {J.~A.}\ \bibnamefont
  {Sanjurjo}}, \bibinfo {author} {\bibfnamefont {C.}~\bibnamefont {Rettori}},
  \bibinfo {author} {\bibfnamefont {I.}~\bibnamefont {Torriani}}, \bibinfo
  {author} {\bibfnamefont {F.}~\bibnamefont {Prado}}, \bibinfo {author}
  {\bibfnamefont {R.~D.}\ \bibnamefont {S\'anchez}}, \bibinfo {author}
  {\bibfnamefont {A.}~\bibnamefont {Caneiro}}, \ and\ \bibinfo {author}
  {\bibfnamefont {S.~B.}\ \bibnamefont {Oseroff}},\ }\bibfield  {title}
  {\enquote {\bibinfo {title} {{Magnetic ordering effects in the Raman spectra
  of
  ${\mathrm{La}}_{1\ensuremath{-}x}{\mathrm{Mn}}_{1\ensuremath{-}x}{\mathrm{O}}_{3}$}},}\
  }\href {\doibase 10.1103/PhysRevB.60.11879} {\bibfield  {journal} {\bibinfo
  {journal} {Phys. Rev. B}\ }\textbf {\bibinfo {volume} {60}},\ \bibinfo
  {pages} {11879--11882} (\bibinfo {year} {1999})}\BibitemShut {NoStop}%
\bibitem [{\citenamefont {Hu}\ \emph {et~al.}(2019)\citenamefont {Hu},
  \citenamefont {Yang}, \citenamefont {Wu}, \citenamefont {Wu}, \citenamefont
  {Zhao}, \citenamefont {Sun}, \citenamefont {Wang}, \citenamefont {He},
  \citenamefont {He}, \citenamefont {Zhang}, \citenamefont {Huang},
  \citenamefont {Li}, \citenamefont {Shi},\ and\ \citenamefont
  {Zhao}}]{Hu_2019PhysRevB}%
  \BibitemOpen
  \bibfield  {author} {\bibinfo {author} {\bibfnamefont {L.~L.}\ \bibnamefont
  {Hu}}, \bibinfo {author} {\bibfnamefont {M.}~\bibnamefont {Yang}}, \bibinfo
  {author} {\bibfnamefont {Y.~L.}\ \bibnamefont {Wu}}, \bibinfo {author}
  {\bibfnamefont {Q.}~\bibnamefont {Wu}}, \bibinfo {author} {\bibfnamefont
  {H.}~\bibnamefont {Zhao}}, \bibinfo {author} {\bibfnamefont {F.}~\bibnamefont
  {Sun}}, \bibinfo {author} {\bibfnamefont {W.}~\bibnamefont {Wang}}, \bibinfo
  {author} {\bibfnamefont {Rui}\ \bibnamefont {He}}, \bibinfo {author}
  {\bibfnamefont {S.~L.}\ \bibnamefont {He}}, \bibinfo {author} {\bibfnamefont
  {H.}~\bibnamefont {Zhang}}, \bibinfo {author} {\bibfnamefont {R.~J.}\
  \bibnamefont {Huang}}, \bibinfo {author} {\bibfnamefont {L.~F.}\ \bibnamefont
  {Li}}, \bibinfo {author} {\bibfnamefont {Y.~G.}\ \bibnamefont {Shi}}, \ and\
  \bibinfo {author} {\bibfnamefont {Jimin}\ \bibnamefont {Zhao}},\ }\bibfield
  {title} {\enquote {\bibinfo {title} {{Strong pseudospin-lattice coupling in
  ${\mathrm{Sr}}_{3}{\mathrm{Ir}}_{2}{\mathrm{O}}_{7}$: Coherent phonon anomaly
  and negative thermal expansion}},}\ }\href {\doibase
  10.1103/PhysRevB.99.094307} {\bibfield  {journal} {\bibinfo  {journal} {Phys.
  Rev. B}\ }\textbf {\bibinfo {volume} {99}},\ \bibinfo {pages} {094307}
  (\bibinfo {year} {2019})}\BibitemShut {NoStop}%
\bibitem [{\citenamefont {Kant}\ \emph {et~al.}(2009)\citenamefont {Kant},
  \citenamefont {Deisenhofer}, \citenamefont {Rudolf}, \citenamefont {Mayr},
  \citenamefont {Schrettle}, \citenamefont {Loidl}, \citenamefont {Gnezdilov},
  \citenamefont {Wulferding}, \citenamefont {Lemmens},\ and\ \citenamefont
  {Tsurkan}}]{Kant_2009PhysRevB}%
  \BibitemOpen
  \bibfield  {author} {\bibinfo {author} {\bibfnamefont {Ch.}\ \bibnamefont
  {Kant}}, \bibinfo {author} {\bibfnamefont {J.}~\bibnamefont {Deisenhofer}},
  \bibinfo {author} {\bibfnamefont {T.}~\bibnamefont {Rudolf}}, \bibinfo
  {author} {\bibfnamefont {F.}~\bibnamefont {Mayr}}, \bibinfo {author}
  {\bibfnamefont {F.}~\bibnamefont {Schrettle}}, \bibinfo {author}
  {\bibfnamefont {A.}~\bibnamefont {Loidl}}, \bibinfo {author} {\bibfnamefont
  {V.}~\bibnamefont {Gnezdilov}}, \bibinfo {author} {\bibfnamefont
  {D.}~\bibnamefont {Wulferding}}, \bibinfo {author} {\bibfnamefont
  {P.}~\bibnamefont {Lemmens}}, \ and\ \bibinfo {author} {\bibfnamefont
  {V.}~\bibnamefont {Tsurkan}},\ }\bibfield  {title} {\enquote {\bibinfo
  {title} {{Optical phonons, spin correlations, and spin-phonon coupling in the
  frustrated pyrochlore magnets ${\text{CdCr}}_{2}{\text{O}}_{4}$ and
  ${\text{ZnCr}}_{2}{\text{O}}_{4}$}},}\ }\href {\doibase
  10.1103/PhysRevB.80.214417} {\bibfield  {journal} {\bibinfo  {journal} {Phys.
  Rev. B}\ }\textbf {\bibinfo {volume} {80}},\ \bibinfo {pages} {214417}
  (\bibinfo {year} {2009})}\BibitemShut {NoStop}%
\bibitem [{\citenamefont {Penc}\ \emph {et~al.}(2004)\citenamefont {Penc},
  \citenamefont {Shannon},\ and\ \citenamefont
  {Shiba}}]{Penc_2004_PhysRevLett}%
  \BibitemOpen
  \bibfield  {author} {\bibinfo {author} {\bibfnamefont {Karlo}\ \bibnamefont
  {Penc}}, \bibinfo {author} {\bibfnamefont {Nic}\ \bibnamefont {Shannon}}, \
  and\ \bibinfo {author} {\bibfnamefont {Hiroyuki}\ \bibnamefont {Shiba}},\
  }\bibfield  {title} {\enquote {\bibinfo {title} {{Half-Magnetization Plateau
  Stabilized by Structural Distortion in the Antiferromagnetic Heisenberg Model
  on a Pyrochlore Lattice}},}\ }\href {\doibase 10.1103/PhysRevLett.93.197203}
  {\bibfield  {journal} {\bibinfo  {journal} {Phys. Rev. Lett.}\ }\textbf
  {\bibinfo {volume} {93}},\ \bibinfo {pages} {197203} (\bibinfo {year}
  {2004})}\BibitemShut {NoStop}%
\bibitem [{\citenamefont {Bergman}\ \emph {et~al.}(2006)\citenamefont
  {Bergman}, \citenamefont {Shindou}, \citenamefont {Fiete},\ and\
  \citenamefont {Balents}}]{Bergman_2006_PhysRevB}%
  \BibitemOpen
  \bibfield  {author} {\bibinfo {author} {\bibfnamefont {Doron~L.}\
  \bibnamefont {Bergman}}, \bibinfo {author} {\bibfnamefont {Ryuichi}\
  \bibnamefont {Shindou}}, \bibinfo {author} {\bibfnamefont {Gregory~A.}\
  \bibnamefont {Fiete}}, \ and\ \bibinfo {author} {\bibfnamefont {Leon}\
  \bibnamefont {Balents}},\ }\bibfield  {title} {\enquote {\bibinfo {title}
  {Models of degeneracy breaking in pyrochlore antiferromagnets},}\ }\href
  {\doibase 10.1103/PhysRevB.74.134409} {\bibfield  {journal} {\bibinfo
  {journal} {Phys. Rev. B}\ }\textbf {\bibinfo {volume} {74}},\ \bibinfo
  {pages} {134409} (\bibinfo {year} {2006})}\BibitemShut {NoStop}%
\bibitem [{\citenamefont {Wang}\ and\ \citenamefont
  {Vishwanath}(2008)}]{Wang_2008_PhysRevLett}%
  \BibitemOpen
  \bibfield  {author} {\bibinfo {author} {\bibfnamefont {Fa}~\bibnamefont
  {Wang}}\ and\ \bibinfo {author} {\bibfnamefont {Ashvin}\ \bibnamefont
  {Vishwanath}},\ }\bibfield  {title} {\enquote {\bibinfo {title} {{Spin Phonon
  Induced Collinear Order and Magnetization Plateaus in Triangular and Kagome
  Antiferromagnets: Applications to ${\mathrm{CuFeO}}_{2}$}},}\ }\href
  {\doibase 10.1103/PhysRevLett.100.077201} {\bibfield  {journal} {\bibinfo
  {journal} {Phys. Rev. Lett.}\ }\textbf {\bibinfo {volume} {100}},\ \bibinfo
  {pages} {077201} (\bibinfo {year} {2008})}\BibitemShut {NoStop}%
\bibitem [{\citenamefont {G\'omez~Albarrac\'{\i}n}\ \emph
  {et~al.}(2013)\citenamefont {G\'omez~Albarrac\'{\i}n}, \citenamefont {Cabra},
  \citenamefont {Rosales},\ and\ \citenamefont
  {Rossini}}]{Albarrac_2013_PhysRevB}%
  \BibitemOpen
  \bibfield  {author} {\bibinfo {author} {\bibfnamefont {F.~A.}\ \bibnamefont
  {G\'omez~Albarrac\'{\i}n}}, \bibinfo {author} {\bibfnamefont {D.~C.}\
  \bibnamefont {Cabra}}, \bibinfo {author} {\bibfnamefont {H.~D.}\ \bibnamefont
  {Rosales}}, \ and\ \bibinfo {author} {\bibfnamefont {G.~L.}\ \bibnamefont
  {Rossini}},\ }\bibfield  {title} {\enquote {\bibinfo {title} {Spin-phonon
  induced magnetic order in the kagome ice},}\ }\href {\doibase
  10.1103/PhysRevB.88.184421} {\bibfield  {journal} {\bibinfo  {journal} {Phys.
  Rev. B}\ }\textbf {\bibinfo {volume} {88}},\ \bibinfo {pages} {184421}
  (\bibinfo {year} {2013})}\BibitemShut {NoStop}%
\bibitem [{\citenamefont {Albarracín}\ \emph {et~al.}(2014)\citenamefont
  {Albarracín}, \citenamefont {Cabra}, \citenamefont {Rosales},\ and\
  \citenamefont {Rossini}}]{Albarracin_2014JPC}%
  \BibitemOpen
  \bibfield  {author} {\bibinfo {author} {\bibfnamefont {F~A~Gómez}\
  \bibnamefont {Albarracín}}, \bibinfo {author} {\bibfnamefont {D~C}\
  \bibnamefont {Cabra}}, \bibinfo {author} {\bibfnamefont {H~D}\ \bibnamefont
  {Rosales}}, \ and\ \bibinfo {author} {\bibfnamefont {G~L}\ \bibnamefont
  {Rossini}},\ }\bibfield  {title} {\enquote {\bibinfo {title} {Spin-phonon
  induced magnetic order in magnetized spin ice systems},}\ }\href {\doibase
  10.1088/1742-6596/568/4/042007} {\bibfield  {journal} {\bibinfo  {journal}
  {Journal of Physics: Conference Series}\ }\textbf {\bibinfo {volume} {568}},\
  \bibinfo {pages} {042007} (\bibinfo {year} {2014})}\BibitemShut {NoStop}%
\bibitem [{\citenamefont {Aoyama}\ and\ \citenamefont
  {Kawamura}(2016)}]{Aoyama_2016_PhysRevLett}%
  \BibitemOpen
  \bibfield  {author} {\bibinfo {author} {\bibfnamefont {Kazushi}\ \bibnamefont
  {Aoyama}}\ and\ \bibinfo {author} {\bibfnamefont {Hikaru}\ \bibnamefont
  {Kawamura}},\ }\bibfield  {title} {\enquote {\bibinfo {title}
  {Spin-lattice-coupled order in heisenberg antiferromagnets on the pyrochlore
  lattice},}\ }\href {\doibase 10.1103/PhysRevLett.116.257201} {\bibfield
  {journal} {\bibinfo  {journal} {Phys. Rev. Lett.}\ }\textbf {\bibinfo
  {volume} {116}},\ \bibinfo {pages} {257201} (\bibinfo {year}
  {2016})}\BibitemShut {NoStop}%
\bibitem [{\citenamefont {Pili}\ and\ \citenamefont
  {Grigera}(2019)}]{Pili_2019_PhysRevB}%
  \BibitemOpen
  \bibfield  {author} {\bibinfo {author} {\bibfnamefont {L.}~\bibnamefont
  {Pili}}\ and\ \bibinfo {author} {\bibfnamefont {S.~A.}\ \bibnamefont
  {Grigera}},\ }\bibfield  {title} {\enquote {\bibinfo {title}
  {{Two-dimensional Ising model with Einstein site phonons}},}\ }\href
  {\doibase 10.1103/PhysRevB.99.144421} {\bibfield  {journal} {\bibinfo
  {journal} {Phys. Rev. B}\ }\textbf {\bibinfo {volume} {99}},\ \bibinfo
  {pages} {144421} (\bibinfo {year} {2019})}\BibitemShut {NoStop}%
\bibitem [{\citenamefont {Gen}\ and\ \citenamefont
  {Suwa}(2022)}]{Gen_2022_PhysRevB}%
  \BibitemOpen
  \bibfield  {author} {\bibinfo {author} {\bibfnamefont {Masaki}\ \bibnamefont
  {Gen}}\ and\ \bibinfo {author} {\bibfnamefont {Hidemaro}\ \bibnamefont
  {Suwa}},\ }\bibfield  {title} {\enquote {\bibinfo {title} {{Nematicity and
  fractional magnetization plateaus induced by spin-lattice coupling in the
  classical kagome-lattice Heisenberg antiferromagnet}},}\ }\href {\doibase
  10.1103/PhysRevB.105.174424} {\bibfield  {journal} {\bibinfo  {journal}
  {Phys. Rev. B}\ }\textbf {\bibinfo {volume} {105}},\ \bibinfo {pages}
  {174424} (\bibinfo {year} {2022})}\BibitemShut {NoStop}%
\bibitem [{\citenamefont {Gao}(2024)}]{Gao_2024_PhysRevB}%
  \BibitemOpen
  \bibfield  {author} {\bibinfo {author} {\bibfnamefont {Shang}\ \bibnamefont
  {Gao}},\ }\bibfield  {title} {\enquote {\bibinfo {title} {{Dynamic
  spin-lattice coupling and statistical interpretation for the molecularlike
  excitations in frustrated pyrochlores}},}\ }\href {\doibase
  10.1103/PhysRevB.110.214420} {\bibfield  {journal} {\bibinfo  {journal}
  {Phys. Rev. B}\ }\textbf {\bibinfo {volume} {110}},\ \bibinfo {pages}
  {214420} (\bibinfo {year} {2024})}\BibitemShut {NoStop}%
\bibitem [{\citenamefont {Ferrari}\ \emph {et~al.}(2021)\citenamefont
  {Ferrari}, \citenamefont {Valent\'{\i}},\ and\ \citenamefont
  {Becca}}]{Ferrari_2021_PhysRevB}%
  \BibitemOpen
  \bibfield  {author} {\bibinfo {author} {\bibfnamefont {Francesco}\
  \bibnamefont {Ferrari}}, \bibinfo {author} {\bibfnamefont {Roser}\
  \bibnamefont {Valent\'{\i}}}, \ and\ \bibinfo {author} {\bibfnamefont
  {Federico}\ \bibnamefont {Becca}},\ }\bibfield  {title} {\enquote {\bibinfo
  {title} {{Effects of spin-phonon coupling in frustrated Heisenberg
  models}},}\ }\href {\doibase 10.1103/PhysRevB.104.035126} {\bibfield
  {journal} {\bibinfo  {journal} {Phys. Rev. B}\ }\textbf {\bibinfo {volume}
  {104}},\ \bibinfo {pages} {035126} (\bibinfo {year} {2021})}\BibitemShut
  {NoStop}%
\bibitem [{\citenamefont {Ferrari}\ \emph {et~al.}(2024)\citenamefont
  {Ferrari}, \citenamefont {Becca},\ and\ \citenamefont
  {Valent\'{\i}}}]{Ferrari_2024_PhysRevB}%
  \BibitemOpen
  \bibfield  {author} {\bibinfo {author} {\bibfnamefont {Francesco}\
  \bibnamefont {Ferrari}}, \bibinfo {author} {\bibfnamefont {Federico}\
  \bibnamefont {Becca}}, \ and\ \bibinfo {author} {\bibfnamefont {Roser}\
  \bibnamefont {Valent\'{\i}}},\ }\bibfield  {title} {\enquote {\bibinfo
  {title} {{Spin-phonon interactions on the kagome lattice: Dirac spin liquid
  versus valence-bond solids}},}\ }\href {\doibase 10.1103/PhysRevB.109.165133}
  {\bibfield  {journal} {\bibinfo  {journal} {Phys. Rev. B}\ }\textbf {\bibinfo
  {volume} {109}},\ \bibinfo {pages} {165133} (\bibinfo {year}
  {2024})}\BibitemShut {NoStop}%
\bibitem [{\citenamefont {Mankovsky}\ \emph {et~al.}(2023)\citenamefont
  {Mankovsky}, \citenamefont {Lange}, \citenamefont {Polesya},\ and\
  \citenamefont {Ebert}}]{Mankovsky_2023_PhysRevB}%
  \BibitemOpen
  \bibfield  {author} {\bibinfo {author} {\bibfnamefont {Sergiy}\ \bibnamefont
  {Mankovsky}}, \bibinfo {author} {\bibfnamefont {Hannah}\ \bibnamefont
  {Lange}}, \bibinfo {author} {\bibfnamefont {Svitlana}\ \bibnamefont
  {Polesya}}, \ and\ \bibinfo {author} {\bibfnamefont {Hubert}\ \bibnamefont
  {Ebert}},\ }\bibfield  {title} {\enquote {\bibinfo {title} {S{pin-lattice
  interaction parameters from first principles: Theory and implementation}},}\
  }\href {\doibase 10.1103/PhysRevB.107.144428} {\bibfield  {journal} {\bibinfo
   {journal} {Phys. Rev. B}\ }\textbf {\bibinfo {volume} {107}},\ \bibinfo
  {pages} {144428} (\bibinfo {year} {2023})}\BibitemShut {NoStop}%
\bibitem [{\citenamefont {Klemens}(1966)}]{Klemens_PhysRev148}%
  \BibitemOpen
  \bibfield  {author} {\bibinfo {author} {\bibfnamefont {P.~G.}\ \bibnamefont
  {Klemens}},\ }\bibfield  {title} {\enquote {\bibinfo {title} {Anharmonic
  decay of optical phonons},}\ }\href {\doibase 10.1103/PhysRev.148.845}
  {\bibfield  {journal} {\bibinfo  {journal} {Phys. Rev.}\ }\textbf {\bibinfo
  {volume} {148}},\ \bibinfo {pages} {845--848} (\bibinfo {year}
  {1966})}\BibitemShut {NoStop}%
\bibitem [{\citenamefont {Men\'endez}\ and\ \citenamefont
  {Cardona}(1984)}]{Cardona_PRB1984}%
  \BibitemOpen
  \bibfield  {author} {\bibinfo {author} {\bibfnamefont {Jos\'e}\ \bibnamefont
  {Men\'endez}}\ and\ \bibinfo {author} {\bibfnamefont {Manuel}\ \bibnamefont
  {Cardona}},\ }\bibfield  {title} {\enquote {\bibinfo {title} {{Temperature
  dependence of the first-order {R}aman scattering by phonons in Si, Ge, and
  $\ensuremath{\alpha}\ensuremath{-}\mathrm{S}\mathrm{n}$: Anharmonic
  effects}},}\ }\href {\doibase 10.1103/PhysRevB.29.2051} {\bibfield  {journal}
  {\bibinfo  {journal} {Phys. Rev. B}\ }\textbf {\bibinfo {volume} {29}},\
  \bibinfo {pages} {2051--2059} (\bibinfo {year} {1984})}\BibitemShut {NoStop}%
\end{thebibliography}

%

\end{document}